\documentclass[preprint,showkeys,secnumarabic,amsfonts,showpacs,amsmath,amssymb]{revtex4}

\usepackage[dvips]{color}
\usepackage{epsfig}
\usepackage{amsmath}
\usepackage{graphicx}
\usepackage{amssymb,epsf}
\usepackage{epstopdf}
\usepackage{makeidx}
\usepackage{verbatim}
\usepackage{hyperref}

\begin{document}
\title{Modified $f(R,G,T)$ Gravity in the Quintom model, and Inflation}

\author{Farzad Milani}\email{f.milani@tvu.ac.ir}\affiliation{Department of Physics, National University of Skills (NUS), Tehran, Iran.}

\date{\today}

\begin{abstract}\label{sec:abstract}
\noindent \hspace{0.35cm}

In this paper, I consider a $f(R, G, T)$ modified gravity model where $R$ represents the Ricci scalar, $G$ denotes the Gauss-Bonnet invariant, and $T$ signifies the trace of the stress-energy tensor. This model is coupled with two distinct types of scalar fields. In the flat Friedmann-Lema\^{\i}tre-Robertson-Walker (FLRW) universe, the necessary conditions for a successful bounce are achieved. Under these circumstances, it is demonstrated that the equation of state (EoS) parameter cannot cross the phantom divider when only the inflaton scalar field is considered. Appropriate conditions to preserve the conservation of energy law are obtained. The absence of radiation domination is confirmed by referencing one of the collections of the Planck 2018 report. Moreover, it has been demonstrated that this is a general model, encompassing other models such as Weyl conformal geometry and the inflation epoch. Numerical calculations and graphs are used to confirm the results.
\end{abstract}



\maketitle

\section{Introduction}\label{sec:Introduction}
Studying the history of the universe's creation has long intrigued philosophers. A key aspect is that humans only grasp a fraction of this history. They are incapable of a complete understanding of its beginning or end. It remains uncertain whether the universe had a definitive commencement with a Big Bang or whether today's universe emerged from a prior one (Bouncing theory \citep{khoury2002big}). Beliefs and non-physical principles influence the acceptance of these theories. The debate on justifying physical issues with non-physical laws and why some emphasize ideological principles in explaining physics, particularly the universe's origin, is a separate topic. Man began formulating fundamental principles to address questions, evolving rules based on these principles. It's crucial to acknowledge that our current knowledge of physics is human-made, subject to replacement by new theories that may either refine or reject existing ones. This fluidity underscores the dynamism of science, anchoring discussions in accepted laws and diminishing superstition's effects over time.

What is evident is that recent observational data clearly show that the universe is accelerating expanding \citep{riess1998observational, perlmutter1999supernova, spergel2003first, spergel2007three}. What is the cause of this accelerated expansion? How long this accelerated expansion will continue? Whether it will one day stop and whether the process of contraction will begin, is a matter for which there is still no convincing answer. Answers to such questions have been updated step by step. For example, to answer for causing the accelerated expansion, an unknown mysterious energy called Dark Energy (DE) was introduced. Today we know DE is the dominant component of the universe, contributing about 
$69\%$ of the total energy in the present day. In this way, the contribution of cold dark matter is $26.1\%$  and just $4.9\%$ is for ordinary (baryonic) matter, and other components such as neutrinos and photons are nearly negligible \citep{Planck2018}.  Plus, if we don't want to return to the first questions raised, these issues can be reviewed from Einstein's General Relativity (GR) theory onward.

GR was not a conformally invariant theory \cite{essen1990general}. GR and quantum mechanics were once considered incompatible. Attempts to merge general relativity and quantum field theory resulted in a theory that seemed to lack meaning. While standard cosmology and quantum field theory do not fully embrace general relativity as a universal gravitational field theory, the issue of quantum gravity is now understood differently than previously believed, rendering these assertions partially incorrect. Some scientists like Feynman quantized general relativity in the early 1960s \citep{Feynman1963Quantum}, while others employed conformal invariant field theories as renormalizable systems \citep{fronsdal1984ghost}.
They utilized unique properties of the gravitational field for linear estimation in GR's equations, resulting in conformal invariance. Consequently, conformal modified gravity was introduced as an alternative theory to GR \citep{faraoni1998conformal, mannheim1997galactic}. The conformal Weyl gravity is the first gravitational theory that maintains scale transformation invariance and is rooted in the local conformal invariance of the metric \citep{takook2010linear}.  

In the standard gravity model, the Einstein-Hilbert action, $\mathcal{S}_{EH}=\frac{1}{16\pi G_r}\int{d^4 x\sqrt{-g}R}$, where $G_r$ is Newton's gravitational constant \cite{Feynman1995Feynman}, by replacing the Ricci scalar $R$, with an arbitrary function of it, we will have a $f(R)$ modified gravity \citep{farajollahi2012cosmological, farajollahi2011cosmic, Bamba2008Future, capozziello2008extended}. If an arbitrary function of the Gauss-Bonnet invariant, $G=R^{2}-4R_{\mu\nu}R^{\mu\nu}+R_{\mu\nu\rho\lambda}R^{\mu\nu\rho\lambda}$, is substituted with the Ricci scalar, $R$, we will have a $f(G)$ modified gravity \citep{sami2005fat, nojiri2005modified, sadeghi2009crossing}. In this way, there are other extensions of the standard gravity model.  If the Ricci scalar, $R$, is replaced by an arbitrary function of the Ricci scalar $R$, and the Gauss-Bonnet invariant $G$, it gives $f(R, G)$ modified gravity \citep{atazadeh2014energy, bamba2010finite, Felice2010cosmological, Felice2010inevitable, Shamir2017energy, Mustafa2020physically, Navo2020stability, bhatti2022analysis}. If it is replaced by an arbitrary function of the Ricci scalar $R$ and the trace of the stress-energy tensor $T$ or the stress-energy tensor of the scalar field $T^{\phi}$ it gives  $f(R, T )$ or $f(R, T^{\phi})$ modified gravity \citep{Harko2011f(RT)gravity, baffou2021inflationary, nashed2023theeffect, singh2016cosmological, rajabi2017unimodular, singh2014reconstruction}, and by replacing an arbitrary function of $G$ and $T$ instead of Ricci scalar, $f(G,T)$ modified gravity will be obtained \citep{sharif2016energy, shamir2018gravastars, shamir2021bouncing}. Finally, if it is replaced by an arbitrary function of all of them together, we will have $f(R, G, T)$ modified gravity that is so popular recently \citep{debnath2020constructions, Chaudhary2023cosmological, ilyas2022energy,  ilyas2024gravastars, ilyas2021compact}.

In section 2, I explore the dynamics of FLRW cosmology in a conformal modified gravity by using the $(- \, , + \, , + \, , +)$ metric signature. Here, I consider a $f(R, G, T)$ modified gravity that is coupled with a single phantom and a single quintessence scalar fields, to explain the effects of dark energy in my model, and introduce an alternative model for the Big Bang theory that is called the bouncing theory. The bouncing theory combines the Big Bang and Big Crunch theories into a unified model  \citep{khoury2002big}. This model proposes a universe that expands from a singularity, collapses back into a singularity, and then repeats the cycle \citep{elitzur2002big}. Therefore, a bouncing universe would involve a continuous pattern of expansion and contraction {\citep{sadeghi2009bouncing,bamba2014bounce,farajollahi2010bouncing,riess2004type,knop2003new,
 riess1999bvri,perlmutter1999measurements,ade2015planck}}. 
 
I often pose a simple question to elaborate on the need for an alternative to the big bang theory. While this question isn't flawless, it serves for basic comprehension. \textit{Does a newborn's creation begin at birth, or was the mother pregnant for a period beforehand?} We typically gauge a person's age from its birth. Yet, it is undeniable that not everyone has undergone a specified pre-birth process, an event whose exact inception might not be pinpointed. This analogy suggests that the universe may have experienced a gestation-like process before its birth. So, given our ongoing efforts in rationalizing the universe's birth, we have overlooked this potential period of cosmic gestation. In this research, I don't want to ignore this process before the big bang. Therefore, I will obtain the suitable conditions for a successful bounce in sections 2, 3, and 6. 

As I said before, in recent years, scientists have made significant progress in understanding dark energy, the force that is thought to drive the universe's accelerated expansion. Various methods have been used to study and elucidate dark energy's nature, shedding light on this enigmatic phenomenon in modern cosmology. 

Key methods used to study dark energy include observational techniques like supernova Ia surveys \citep{smecker1991type, ruiz1995type, nomoto1997type}, Baryon Acoustic Oscillations (BAO) \citep{eisenstein2005detection}, and Cosmic Microwave Background (CMB) {\cite{spergel2003first,spergel2007three}} radiation measurements. These methods have yielded valuable data that helps constrain dark energy's properties and understand its effects on the universe.

In light of the compelling evidence regarding the accelerated expansion of the universe, theoretical frameworks of bouncing theory must adapt to include mathematical models that elucidate the intricate relationship between dark energy and dark matter. One potential scenario involves gradual and seamless fluctuations in the mass-energy distribution of dark energy components over time  {\cite{astier2006supernova,riess2007new,wood2007observational,bamba2012dark}}. 

Theoretical models, such as the Lambda Cold Dark Matter ($\Lambda$CDM) or the cosmological constant \citep{padmanabhan2003cosmological, sahni2000case}, quintessence \citep{ratra1988cosmological}, phantom \citep{caldwell2002phantom}, quintom \citep{feng2005dark, guo2005cosmological}, tachyon \citep{sen2002tachyon}, k-essence \citep{armendariz2000dynamical}, dilaton \citep{gasperini2001quintessence}, hessence \citep{wei2005hessence}, DBI-essence \citep{gumjudpai2009generalized, martin2008dbi} have also been proposed to explain dark energy. These models present different viewpoints on dark energy's origin and behavior, providing a framework for comprehending its role in the universe's expansion.

By comparing these diverse methods, one can enhance our understanding of dark energy and its implications for the cosmos' future. Each approach offers unique insights into dark energy's nature, and by considering them collectively, a more complete understanding of this mysterious force and its cosmic impact can be developed.

For example, the $\Lambda$CDM model, with an Equation of State (EoS), $\omega=-1$, is strongly backed by the Wilkinson Microwave Anisotropy Probe (WMAP) \cite{bennett2003first,bennett2011seven,bennett20141,bennett2013nine}, making it a key theoretical contender for modeling dark energy. Observations in astronomy reveal that the cosmological constant's numerical value is significantly smaller than what prior theories had predicted {\cite{weinberg1989cosmological}}. Hence, employing a $\Lambda\simeq 0$ approximation in the calculations aligns with Einstein's universe equation. Current EoS parameter values hover around $\omega =-1\pm 0.1$ {\cite{tonry2003cosmological,tegmark2004cosmological,
bennett2003first,bennett2011seven,bennett20141,bennett2013nine,weinberg1989cosmological,
copeland2006dynamics,seljak2005cosmological,tegmark2005does}} , acknowledging that this value might vary with time \citep{ghanaatian2014bouncing, ghanaatian2018bouncing, sadeghi2008non, farajollahi2011stability, farajollahi2010cosmic, farajollahi2011stability1, farajollahi2011stability2}.

For other purposes, I also need to review quintessence and phantom concepts. They are dark energy models that involve scalar fields with unique characteristics and behaviors.

Quintessence is a hypothetical dark energy form linked to a scalar field with dynamic energy density that evolves over time. Its scalar field usually exhibits positive kinetic energy and a potential energy term driving universe expansion acceleration. Its EoS parameter typically ranges between -1 and 0, resulting in moderate cosmic expansion acceleration. Quintessence is viewed as a more conventional dark energy form compared to phantom dark energy.

Phantom dark energy, another hypothetical model, features a scalar field with negative kinetic energy, leading to unique properties. Its potential energy term may exhibit specific characteristics generating negative pressure. The EoS parameter for phantom dark energy is below -1, violating the null energy condition and causing exponential universe expansion, known as the "Big Rip" scenario \citep{ron2003cosmic, schewe2003thebigrip, chimento2004onbigrip}. Phantom dark energy is linked to instabilities and peculiar behaviors challenging conventional physical intuition. 

The quintessence and phantom concepts combination is called the \textbf{\textit{quintom}} dark energy model in theoretical cosmology. The term "quintom" is derived from the words "quintessence" and "phantom," highlighting the hybrid nature of the model.

In the quintom model, the equation of state is equal to a sum of two components: one component that resembles the quintessence field and another that resembles the phantom field. This combination allows the dark energy field and other parameters to interact with each other.

The quintom model is considered one of the top models for describing the bouncing theory. Its equation of state can transition from below -1 to above -1, a characteristic observed in numerous phenomenological models \cite{caldwell2003phantom, carroll2003can, cline2004phantom, mcinnes2002ds, melchiorri2003state, nesseris2004fate, onemli2004quantum, alam2004case, padmanabhan2002accelerated, padmanabhan2002can, hao2003attractor, singh2003cosmological, carroll2005can, aref2006exact,guo2005interacting, boisseau2000reconstruction, sahni2003braneworld, mcinnes2005phantom, perivolaropoulos2005constraints, anisimov2005b, witten1986non, witten1986interacting}. So, Identifying a specific mathematical model that aligns with foundational principles and predicts $\omega$ crossing over the $-1$ separator line, is a key objective.

On the other hand, the cosmic inflation epoch is a crucial phase in the early universe that is believed to have occurred within the first fraction of a second after the Big Bang. Proposed by Alan Guth in the early 1980s, cosmic inflation posits a rapid and exponential expansion of the universe, leading to the smoothing out of inconsistencies in the CMB radiation and the formation of primordial density fluctuations that later grew into the large-scale structures observed today \cite{Guth1981}. This period of inflationary expansion is fundamental to our understanding of the universe's evolution and structure, as it provides a solution to several long-standing problems in cosmology, such as the horizon problem and the flatness problem \cite{Linde2008}.

Recent advancements in observational cosmology, particularly from experiments such as the CMB observations by the Planck satellite, have provided strong evidence in support of the inflationary paradigm \cite{Planck2018}. By studying the statistical properties of the CMB temperature fluctuations and the polarization of the CMB, researchers have been able to test specific predictions of inflationary models and constrain their parameters \cite{Peiris2003}.

In section 2, the wave equations of motion and the $00$ and $ii$ components of the energy-momentum tensor of the model are used to derive the modified Friedmann equations. This leads to obtaining the energy density, and the pressure of the dark energy. The analysis focuses on the conditions in which to achieve a successful bounce,  $\omega$ crosses over $-1$. Additionally, the study delves into the requirements for ensuring the conservation of energy law in a general case. In Section 3, with the recent observational data from the Planck 2018 reports \citep{Planck2018}, the energy density of total matter, baryonic matter, cold dark matter, radiation, and dark energy at the present time are estimated. In section 4, Weyl conformal geometry is the first example covered by the model. In section 5, examining the inflationary epoch for a single inflaton scalar field is the second example analyzed in my model. The numerical solutions for a successful bounce are discussed in section 6, followed by a conclusion of my discussion in section 7.

\section{The Model}\label{Model}

In the modified gravity, one can replace the Ricci scalar term of the Einstein-Hilbert Lagrangian density by an arbitrary function of the Ricci scalar, $R$, the Gauss-Bonnet invariant, $G=R^{2}-4R_{\mu\nu}R^{\mu\nu}+R_{\mu\nu\rho\lambda}R^{\mu\nu\rho\lambda}$, and the trace of stress-energy tensor of the matter and radiation, $T=g^{\mu\nu}T_{\mu\nu}^{(m,r)}$, which is coupled with a phantom scalar field, $\phi=\phi(t)$, and a standard scalar field, $\psi=\psi(t)$, called \textit{quintessence} if we consider the quintom model, and called $\textit{inflaton}$ if we want to study our model during the period of inflation as, $R\rightarrow \alpha f(R, G, T)+\beta \Xi(\phi,\psi)$. These fields are just dependent on cosmic time, $t$. Typically, $\phi(t)$ introduced the open
string tachyon, and $\psi(t)$ the closed string tachyon. The total Lagrangian density of our model is,
\begin{eqnarray}
\mathcal{L}=\frac{1}{2k^2}f(R, G, T)+\frac{1}{\kappa^2_s}\Xi(\phi,\psi)+\mathcal{L}_{m,r},
\end{eqnarray}
and thus our action is yielded by,
\begin{eqnarray}\label{ac1}
\mathcal{S}=\int{d^4x\sqrt{-g}\left(\frac{f(R, G, T)}{2k^2}+\frac{\Xi(\phi,\psi)}{\kappa^2_s}+\mathcal{L}_{m,r}\right)},
\end{eqnarray}
where $k^2=8\pi G_r/c^4=M^2_s/M^2_p$, $g=\textrm{det}g_{\mu\nu}$, $G_r\cong 6.67430\times 10^{-11} N.m^2/kg^2$ is the universal Gravitational constant that is a fundamental constant of nature, $c\cong 299,792,458\, m/s$ is the light speed in the vacuum, $M_p$ is the reduced Planck mass, $M_s$ is the string mass, $\kappa^2_s = m^2_p M^2_s/M^2_p = m^2_p k^2$ is the dimensionless open string coupling constant, $m_p$ is the dimensionless parameter of reduced Planck mass, and $\mathcal{L}_{m,r}$ is the matter and radiation Lagrangian density. In addition, 
\begin{eqnarray}\label{Xi}
\Xi(\phi,\psi)=\frac{1}{2}g^{\mu\nu}\partial_{\mu}\phi\partial_{\nu}\phi-\frac{1}{2}g^{\mu\nu}\partial_{\mu}\psi\partial_{\nu}\psi-V(\phi,\psi),
\end{eqnarray}
and $V(\phi,\psi)$ is an arbitrary potential function dependent on dimensionless $\phi(t)$ and $\psi(t)$. If we consider the metric of the flat FLRW universe in the Cartesian coordinate,
\begin{eqnarray}\label{FRWmetric}
ds^2=-dt^2+a^2(t)\sum_{i=1}^{3}(dx^i)^2\,,
\end{eqnarray}
where $a = a(t)$ is the scalar factor. The variation of the action (\ref{ac1}) with respect to scalar fields, $\phi$ and $\psi$, provides the wave equation of motion for them,
\begin{eqnarray}
\ddot{\phi}+3H\dot{\phi}-V_{,\phi}&=&0\label{EOM-phi}\\
\ddot{\psi}+3H\dot{\psi}+V_{,\psi}&=&0 \label{EOM-psi}
\end{eqnarray}
where $H=\frac{\dot{a}}{a}$, is the Hubble parameter, the dot insinuates to a cosmic time, $t$ derivation, $V_{,\phi}$ and $V_{,\psi}$ indicate differentiation of $V(\phi,\psi)$ with respect to the $\phi$ and $\psi$, respectively \citep{farajollahi2010cosmic}.

The variation of the action (\ref{ac1}) with respect to the inverse metric $g^{\mu\nu}$, as has been shown in the section \ref{subsec:Modified E-M-T} (Appendix), yields us the energy-momentum tensor as,
\begin{eqnarray}
T_{\mu\nu}^{(R)}+T_{\mu\nu}^{(G)}&=&k^2 \left(T_{\mu\nu}^{(m,r)}+T_{\mu\nu}^{(T)}+\frac{1}{\kappa^2_s} T_{\mu\nu}^{(\Xi)}\right)\label{Energy-Momentum Tensor}
\end{eqnarray} 
where 
\begin{eqnarray}
T_{\mu\nu}^{(R)}&=& f_R R_{\mu\nu}-\frac{1}{2} g_{\mu\nu} f + (g_{\mu\nu}\square - \nabla_\mu\nabla_\nu)f_R, \label{T(R)}\\
T_{\mu\nu}^{(G)}&=&2R\left(f_G R_{\mu\nu}+(g_{\mu\nu}\square - \nabla_\mu\nabla_\nu)f_G\right)
-4\left(f_G R_{\mu}^{\rho}R_{\rho\nu}+(R_{\mu\nu}\square+g_{\mu\nu}R^{\rho\lambda}\nabla_{\rho}\nabla_{\lambda})f_G\right)\nonumber\\
&-&4f_G\left(R_{\mu\rho\nu\lambda}R^{\rho\lambda}-\frac{1}{2}R_{\mu}^{\rho\lambda\xi}R_{\nu\rho\lambda\xi}\right)+4\left(R_{\mu}^{\rho}\nabla_{\nu}\nabla_{\rho}+R_{\nu}^{\rho}\nabla_{\mu}\nabla_{\rho}+R_{\mu\rho\nu\lambda}\nabla^{\rho}\nabla^{\lambda}\right)f_G,\label{T(G)}\\
T_{\mu\nu}^{(\Xi)}&=&-\partial_{\mu}\phi\partial_{\nu}\phi+\partial_{\mu}\psi\partial_{\nu}\psi
+\frac{1}{2}g_{\mu\nu}\left(g^{\alpha\beta}(\partial_{\alpha}\phi\partial_{\beta}\phi-\partial_{\alpha}\psi\partial_{\beta}\psi)
-2V(\phi,\psi)\right),\label{T(Xi)}\\
T_{\mu\nu}^{(m,r)}&=&-\frac{2}{\sqrt{-g}}\frac{\partial(\sqrt{-g}\mathcal{L}_{m,r})}{\partial g^{\mu\nu}}=g_{\mu\nu}\mathcal{L}_{m,r}-2\frac{\partial\mathcal{L}_{m,r}}{\partial g^{\mu\nu}},\label{T(m)}\\
T_{\mu\nu}^{(T)}&=&-\frac{f_T}{k^2}\left(T_{\mu\nu}^{(m,r)}+\Theta_{\mu\nu}\right),\qquad\text{when}\qquad \Theta_{\mu\nu}=g^{\alpha\beta}\frac{\partial T_{\alpha\beta}^{(m,r)}}{\partial g^{\mu\nu}}\cdot \label{T(T)}
\end{eqnarray}

As a simultaneous description of non-relativistic matter and radiation in the perfect fluid form, one can use 
\begin{eqnarray}
T_{\mu\nu}^{(m,r)} = \left(\rho+ p\right)u_{\mu}u_{\nu} + p g_{\mu\nu},\label{Tmunu}
\end{eqnarray}
where $\rho \doteq \rho_m +\rho_r$ and $p\doteq p_m + p_r$, and $\rho_m$ denotes the density of the non-relativistic matter, when $\rho_r$ is the radiation density. Plus, the four-velocity $u_{\mu}$ satisfies $u_{\mu}u^{\mu}=-1$ and $u^{\mu}\nabla_{\nu}u_{\mu}=0$. If there is no interaction between non-relativistic matter and radiation, then these components in the standard mode, obey separately the conservation laws 
\begin{eqnarray}
\dot{\rho}+3H(\rho+p)=0,\qquad\text{so}\qquad
\dot{\rho}_{m}+3H\rho_{m}=0,\qquad\text{and}\qquad \dot{\rho}_{r}+4H\rho_{r}=0,\label{Continuity-m,r}
\end{eqnarray}
when  $\omega_m=p_m/\rho_m\approx 0$, and $\omega_r=p_r/\rho_r = 1/3$. With using Eq. (\ref{T(m)}), $\Theta_{\mu\nu}$ which has been given at the section \ref{subsec:Theta} (Appendix), is yielded by
\begin{eqnarray}
\Theta_{\mu\nu}= g_{\mu\nu}\mathcal{L}_{m,r}-2T_{\mu\nu}^{(m,r)}-2g^{\alpha\beta}\frac{\partial^2\mathcal{L}_{m,r}}{\partial g^{\alpha\beta}\partial g^{\mu\nu}}\cdot\label{Thetamunu}
\end{eqnarray}
Now, by using $\mathcal{L}_{m,r} = p_m+p_r=p$, the Eq. (\ref{Thetamunu}) summarizes to
\begin{eqnarray}
\Theta_{\mu\nu}= -2T_{\mu\nu}^{(m,r)}+p g_{\mu\nu}.\label{Theta}
\end{eqnarray}
So, for FLRW metric, one can yield 
\begin{eqnarray}
T &=& -\rho+3p\,=-(1-3\omega)\rho\approx -\rho_m,\label{T}\\
\Theta &=& 2\left(\rho -p\right)=\,\,2(1-\omega)\rho\,\,\approx 2\left(\rho_m+\frac{2}{3}\rho_r\right)\label{Theta},
\end{eqnarray}
where $p=\omega{\rho}$, $\Theta=\Theta_{\mu\nu}g^{\mu\nu}$, and
\begin{eqnarray}
T+\Theta=2\left(\rho +p\right)=2\rho(1+\omega) \approx 2\left(\rho_m+\frac{4}{3}\rho_r\right) \cdot \label{T+Theta}
\end{eqnarray}

The $00$ and $ii$ components of the Eq. (\ref{Energy-Momentum Tensor}), show the energy density and pressure respectively by
\begin{eqnarray}
\rho_R+\rho_G&=&k^2\left(\rho+\rho_T\right)+\frac{1}{m^2_p}\rho_{\Xi},\label{f1}\\
p_R+p_G&=& k^2 \left(p+p_T\right)+\frac{1}{m^2_p}p_{\Xi}.\label{f2}
\end{eqnarray}
where
\begin{eqnarray}
\rho_R&=&-3H^2 f_R + 3H\dot{f}_R- 3 \dot{H}f_R + \frac{f}{2}\label{rho_R}\\
p_R&=& \,\,\,\,3H^2f_R- 2H\dot{f}_R + \,\,\,\dot{H}f_R - \frac{f}{2}- \ddot{f}_R\label{p_R}\\
\rho_G&=&-12H^{2}(H^{2}+\dot{H})f_{G}+H(3H^{2}-9\dot{H})\dot{f}_{G}-9(H^2+\dot{H})\ddot{f}_{G},\label{rho_G}\\
p_G&=&\,\,\,\,\,12H^{2}(H^{2}+\dot{H})f_{G}+H(7H^{2}-5\dot{H})\dot{f}_{G}-(H^2-3\dot{H})\ddot{f}_{G},\label{p_G}\\
\rho_T&=& \frac{f_{T}}{k^2}\left(\rho +p\right) \approx \frac{f_{T}}{k^2}\left(\rho_{m}+\frac{4}{3}\rho_{r}\right),\label{rho_T}\\
p_T&=&0,\label{p_T}\\
\rho_{\Xi}&=&-\frac{\dot{\phi}^2}{2}+\frac{\dot{\psi}^2}{2}+V(\phi,\psi),\label{rho_Xi}\\
p_{\Xi}&=&-\frac{\dot{\phi}^2}{2}+\frac{\dot{\psi}^2}{2}-V(\phi,\psi),\label{p_Xi}
\end{eqnarray}
when $ R=6\dot{H}+12H^2$, and $G=24H^2(\dot{H}+H^2)$.

As I have shown in section \ref{subsec:Covariant} (Appendix), by taking the covariant divergence of Eq. (\ref{Energy-Momentum Tensor}) we have
\begin{eqnarray}
\nabla^{\mu}T_{\mu\nu}^{(R)}&=& -\frac{1}{2}g_{\mu\nu}f_G\nabla^{\mu}G-\frac{1}{2}g_{\mu\nu}f_T\nabla^{\mu}T,\label{Co_TR}\\
\nabla^{\mu}T_{\mu\nu}^{(\Xi)} & = & 0\cdot\label{Co_TXi} 
\end{eqnarray}
So, for the covariant divergence of $T_{\mu\nu}^{(m,r)}$, one can yielded
\begin{eqnarray}
\nabla^{\mu} \left(k^2 T_{\mu\nu}^{(m,r)}-f_{T}\left(T_{\mu\nu}^{(m,r)}+\Theta_{\mu\nu}\right)-T_{\mu\nu}^{(G)}\right)+\frac{1}{2}g_{\mu\nu}f_G\nabla^{\mu}G+\frac{1}{2}g_{\mu\nu}f_T\nabla^{\mu}T= 0,\label{Co_Ttotal}
\end{eqnarray}
therefore, 
\begin{eqnarray}
\nabla^{\mu}T_{\mu\nu}^{(m,r)}&=&\frac{f_T}{k^2-f_T}\left(\left(T_{\mu\nu}^{(m,r)}+\Theta_{\mu\nu}\right)\nabla^{\mu}\ln(f_T)+\nabla^{\mu}\Theta_{\mu\nu}-\frac{1}{2}g_{\mu\nu}\nabla^{\mu}T\right)\nonumber\\
&+&\frac{1}{k^2-f_T}\left(\nabla^{\mu}T_{\mu\nu}^{(G)}-\frac{1}{2}g_{\mu\nu}f_G\nabla^{\mu}G\right)\cdot\label{Co_Tm}
\end{eqnarray}
Again, by using $ R=6\dot{H}+12H^2$, $G=24H^2(\dot{H}+H^2)$, the covariant divergence of $T_{\mu\nu}^{(R)}$ and $T_{\mu\nu}^{(G)}$, can be rewritten as
\begin{eqnarray}
\nabla^{\mu}T_{\mu\nu}^{(R)}&=& 12f_R H\dot{H} + 3f_R\ddot{H} - \frac{\dot{f}}{2},\label{Co_TR_H}\\
\nabla^{\mu}T_{\mu\nu}^{(G)}&=&9(H^2+\dot{H})\dddot{f}_G+9(3H^3+5H\dot{H}+\ddot{H})\ddot{f}_G+3(4H^2f_G+3H\dot{f}_G)\ddot{H}\nonumber\\
&+&3(8Hf_G+3\dot{f}_G)\dot{H}^2+3(16H^3f_G+15H^2\dot{f}_G)\dot{H}-18H^4\dot{f}_G,\label{Co_TG_H} 
\end{eqnarray}
and consequently,
\begin{eqnarray}
\nabla^{\mu}T_{\mu\nu}^{(G)}-\frac{1}{2}g_{\mu\nu}f_G\nabla^{\mu}G &=&9(H^2+\dot{H})\dddot{f}_G+9(3H^3+5H\dot{H}+\ddot{H})\ddot{f}_G\nonumber\\
&-&9\dot{f}_G(2H^4-5H^2\dot{H}-H\ddot{H}-\dot{H}^2)\label{Co_TG_fG_H} 
\end{eqnarray}
Now, by replacing Eqs. (\ref{Tmunu}), (\ref{Thetamunu}), (\ref{T}), and (\ref{Co_TG_fG_H}) in Eq. (\ref{Co_Tm}) one can conclude 
\begin{eqnarray}
\dot{\rho}+3H(\rho+p)&=&\frac{9}{k^2-f_T}\left( 2H^4-5H^2\dot{H}-H\ddot{H}-\dot{H}^2\right)\dot{f}_G\nonumber\\
&-&\frac{9}{k^2-f_T}\left((H^2+\dot{H})\dddot{f}_G+(3H^3+5H\dot{H}+\ddot{H})\ddot{f}_G\right)\nonumber\\
&-&\frac{1}{k^2-f_T}\left(\left(\rho+p\right)\dot{f}_T+6\left(\left(\rho+p\right)H+\frac{5\dot{\rho}}{12}-\frac{\dot{p}}{12}\right)f_T\right)
\end{eqnarray}
So, comparing with the standard perfect fluid form Eq.(\ref{Continuity-m,r}), one can consequent the model conditionally supports the conservation of energy law if
\begin{eqnarray}
& &\left(\rho+p\right)\dot{f}_T+\left( 6\left(\rho+p\right)H+\frac{5\dot{\rho}}{2}-\frac{\dot{p}}{2}\right)f_T =\nonumber\\
&9&\left(( 2H^4-5H^2\dot{H}-H\ddot{H}-\dot{H}^2)\dot{f}_G-(3H^3+5H\dot{H}+\ddot{H})\ddot{f}_G-(H^2+\dot{H})\dddot{f}_G\right)\cdot \label{conserve-condition}
\end{eqnarray}

On the other hand, the Einstein Field Equation (EFE), $G_{\mu\nu}+\Lambda g_{\mu\nu}=R_{\mu\nu}-\frac{1}{2}g_{\mu\nu}R+\Lambda g_{\mu\nu}=k^2 {T}_{\mu\nu}$, in the Einstein-Hilbert form, gives the standard modified Friedmann equations. The left-hand side describes the geometry of space-time by the Einstein tensor $G_{\mu\nu}$ that in the most general form, a cosmological constant $\Lambda g_{\mu\nu}$ may be added to, and the right-hand side describes the total matter distribution by the total modified energy-momentum tensor, ${T}_{\mu\nu}$:
\begin{eqnarray}
3H^2 -\Lambda&=&k^2\left(\rho +\rho_T\right),\label{f1}\\
-2\dot{H}-3H^2 +\Lambda &=& k^2 p.\label{f2}
\end{eqnarray}

When the cosmological constant, $\Lambda$, is zero, EFE reduces to the field equation of general relativity. When $T_{\mu\nu}$ is zero, the EFE describes empty space (a vacuum). $\Lambda$ has the same effect as an intrinsic energy density of the vacuum, $\rho_{vac}$, so $\Lambda_{vac} \equiv k^2 \rho_{vac}$. 

In the $\Lambda$CMD model, dark energy is represented in the vacuum through a distinct cosmological constant, $\Lambda_{vac}$. Yet, in other alternative models such as the quintom model, one might consider $\Lambda$ as equivalent to $\rho_{vac}$, a concept that varies with time. Suppose these two concepts precisely equal (i.e., $\Lambda =\Lambda_{vac}$), in that case, it can be stated that the density of dark energy and the density of the vacuum are equivalent (i.e., $\rho_{DE}=\rho_{vac}$). So, by using $\Lambda\equiv k^2 \rho_{DE}$, the equations (\ref{f1}), and (\ref{f2}) can be rewritten as,
\begin{eqnarray}
3H^2 -k^2\rho_{DE}&=&k^2\left(\rho +\rho_T\right),\label{f11}\\
-2\dot{H}-3H^2 +k^2\rho_{DE} &=& k^2 p,\label{f22}
\end{eqnarray}
where
\begin{eqnarray}
k^2 \rho_{DE}&=&3H^2-(\rho_R+\rho_G)+\frac{1}{m^2_p}\rho_{\Xi}\nonumber\\
&=&3H^2(1+ f_R) - 3H\dot{f}_R+ 3 \dot{H}f_R - \frac{f}{2}\nonumber\\
&+&9(H^2+\dot{H})\ddot{f}_G+3H(3\dot{H}-H^2)\dot{f}_G+12H^2(\dot{H}+H^2)f_G+\frac{\rho_{\Xi}}{m^2_p}\cdot\label{rho-DE}
\end{eqnarray}
In this way, just in the $\Lambda$CDM model, not in the quintom model, by using $\omega_{DE}=-1$ or $ \rho_{DE}=-p_{DE}$ in Eq. (\ref{f22}), we have 
\begin{eqnarray}
k^2 p_{DE}&=&-2\dot{H}-3H^2-(p_R+p_G)+\frac{1}{m^2_p}p_{\Xi}\nonumber\\
&=&(-2\dot{H}-3H^2)(1+f_R)+ 2H\dot{f}_R +\dot{H}f_R + \frac{f}{2}+ \ddot{f}_R\nonumber\\
&-&12H^{2}(\dot{H}+H^{2})f_{G}+H(5\dot{H}-7H^{2})\dot{f}_{G}+(H^{2}-3\dot{H})\ddot{f}_{G}+\frac{p_{\Xi}}{m^2_p}\cdot\label{p-DE}
\end{eqnarray}
So, 
\begin{eqnarray}
\rho_{eff}&=&\rho\left(1+\frac{f_T}{k^2}\right) +p\frac{f_T}{k^2} + \rho_{DE}=\rho\left(1+\frac{f_T}{k^2}(1+\omega)\right)+\rho_{DE}=\frac{3H^2}{k^2},\label{f1-E}\\
p_{eff}&=& p-\rho_{DE}=-\frac{2\dot{H}+3H^2}{k^2} \cdot\label{f2-E}
\end{eqnarray}
For their effective Equation of State (EoS) parameter, $\omega_{eff}=\frac{p_{eff}}{\rho_{eff}}$, we will have,
\begin{eqnarray}\label{omega_eff}
\omega_{eff}&=&-1-\frac{2}{3}\frac{\dot{H}}{H^2}\nonumber\\
&\approx&-1+\frac{\rho_{m}+\frac{4}{3}\rho_{r}\left(1+\frac{f_T}{k^2}\right)}
{\rho\left(1+\frac{f_T}{k^2}\right)+p\frac{f_T}{k^2}+\rho_{DE}}\cdot
\end{eqnarray}

From the first part of equation (\ref{omega_eff}), we find $\omega_{eff} <-1$ for the phantom if  $\dot{H} > 0$, against $\omega_{eff} >-1$ for the quintessence if $\dot{H} < 0$, respectively. To consider the cosmological evolution of the effective EoS parameter, $\omega_{eff}$, it is possible to indicate some conditions of analytically crossing over the phantom divider line ($\omega_{eff}\rightarrow -1$). To seek
this possibility, the value of $\rho_{eff}\left(1 + \omega_{eff}\right)$ must be disappeared
at the bouncing point although $\frac{d}{dt}(\rho_{eff}+p_{eff})\neq 0$.  So, by using  Eqs. (\ref{f1-E}) and (\ref{f2-E}) we have,
\begin{eqnarray}\label{rho plus p}
\frac{d}{dt}(\rho_{eff}&+&p_{eff})=-2\ddot{H}\neq 0,
\end{eqnarray}

\section{Non-radiation domination in the recent observational results}
The density of non-relativistic matter ($\rho_m$) is  not directly provided in the Planck 2018 report as a standalone value. Instead, the Planck collaboration typically presents the cosmological density parameter for total matter ($\Omega_m=\Omega_b+\Omega_c+\Omega_r$). Here $\Omega_b$, $\Omega_c$, and $\Omega_r$ represent the density parameters for ordinary (baryonic) matter, cold dark matter, and other components like neutrinos and photons (radiation). By considering the critical density ($\rho_{critical}\doteq \rho_{c}$), one can estimate the density of total matter. In this way, the critical density ($\rho_{c}$) is defined as the density required for the universe to be flat (i.e., critical density corresponds to $\Omega =\Omega_{\Lambda}+\Omega_m = 1$). The value of the critical density can be calculated using the formula:
\begin{eqnarray}
\rho_{c} = \frac{3H^{2}_{0}}{k^2 c^2}
\end{eqnarray}

The Planck 2018 report  ($68\%$, TT,TE,EE+lowE+lensing+BAO)  gives $\Omega_{\Lambda} = 0.6897 \pm 0.0057$, $\Omega_m=0.3103 \pm 0.0057$, $\Omega_b h^2=0.02234 \pm 0.00014$, $\Omega_c h^2= 0.11907 \pm 0.00094$, and the current rate of expansion of the universe, the Hubble Constant, 
\begin{eqnarray}
H_0 &=& 100h \,\text{km}\,\text{s}^{-1}\,\text{Mpc}^{-1}=67.66 \pm 0.42 \,\text{km}\,\text{s}^{-1}\,\text{Mpc}^{-1},
\end{eqnarray}
where $h$ is a dimensionless number parameterizing of ignorance \citep{Planck2018}, and in the result $\Omega_b=0.0488 \pm 0.0003$, and $\Omega_c= 0.2601 \pm 0.002$. 
Once we have the critical density, we can estimate the cosmological constant, $\Lambda_0$, the density of total matter, $\rho_{_{0}m}$, baryonic matter, $\rho_{_{0}b}$, cold dark matter, $\rho_{_{0}c}$, radiation, $\rho_{_{0}r}$, and the density of dark energy, $\rho_{_{0}DE}$, at the present time as
\begin{eqnarray}
\Lambda_{0} &\cong & 1.1069 \times 10^{-52}m^{-2}\\
\rho_{_{0}m} &=& \rho_{c}\Omega_m\cong 2.3981\times 10^{-10}Pa \\
\rho_{_{0}b} &=& \rho_{c}\Omega_b\cong 0.3771\times 10^{-10}Pa \\
\rho_{_{0}c} &=& \rho_{c}\Omega_c\cong 2.0101\times 10^{-10}Pa \\
\rho_{_{0}r} &=& \rho_{c}\Omega_r\cong 2.8167\times 10^{-13}Pa \\
\rho_{_{0}DE}&=&\rho_c\Omega_{\Lambda}\cong 5.3302\times 10^{-10}Pa\label{H0}
\end{eqnarray}
when $Pa=J/m^3$ is the Pascal unit. So, the contribution of radiation is negligible against other parameters, and in the absence of radiation, using eqs. (\ref{f1-E}), (\ref{f2-E}), and (\ref{omega_eff}), for a \textbf{time-independent} effective equation of state, $\omega_i$, one can get 
\begin{eqnarray}
a \varpropto \rho_{eff}^{3(1+\omega_{i})},\qquad\text{and}\qquad a \varpropto t^{\frac{2}{3(1+\omega_{i})}}.
\end{eqnarray} 
These equations show us the universe's effective energy density determines its evolution. 

$\bullet$ When $\omega_{i}>0$, it signifies a type of energy known as "stiff matter" \citep{Vilenkin1983} with a high energy density and pressure. In this case, if $a \varpropto\rho_{eff}^{\frac{2}{n}}$ and $a \varpropto t^n$, then $0<n<1$.  In fact, the concept of "stiff matter" in cosmology refers to a form of matter with an equation of state parameter $\omega_i=1$, where the pressure is equal to the energy density. So,  $a \varpropto\rho_{eff}^{6}$ and $a \varpropto t^{\frac{1}{3}}$. Stiff matter has interesting properties, including providing a counterexample to the cosmic no-hair theorem. Stiff matter behaves differently from ordinary matter or radiation. This kind of energy can result in rapid expansion of the universe and can have significant implications in cosmological scenarios.

$\bullet$ For the case where $\omega_{i}=0$, this implies a form of energy often referred to as "non-relativistic matter". In this case, $a \varpropto\rho_{eff}^{3}$ and $a \varpropto t^{\frac{2}{3}}$. It typically represents matter with negligible pressure such as dark matter in the universe. Non-relativistic matter plays a crucial role in the dynamics of the cosmos, influencing the formation of structures like galaxies and galaxy clusters.

$\bullet$ When $-\frac{1}{3}<\omega_{i}<0$, we are dealing with energy that corresponds to a substance known as "relativistic matter". In this case, $a \varpropto\rho_{eff}^{\frac{2}{n}}$ and $a \varpropto t^n$, and $\frac{2}{3}<n<1$.  Relativistic matter is often associated with radiation or ultra-relativistic particles. This type of energy tends to slow down the expansion of the universe due to its negative pressure.

$\bullet$ When $\omega_{i}=-\frac{1}{3}$, it points to a special case where the energy behaves similarly to that of radiation. In this case, $a \varpropto\rho_{eff}^{2}$ and $a \varpropto t$.  The equation of state parameter being precisely $-\frac{1}{3}$ corresponds to a form of matter where energy density scales inversely with the volume of space, leading to a specific rate of cooling as the universe expands.

$\bullet$ When $-1<\omega_{i}<-\frac{1}{3}$, we are dealing with a form of energy that falls under the category of "phantom energy". In this case, $a \varpropto\rho_{eff}^{\frac{2}{n}}$ and $a \varpropto t^n$, and $n>1$.  This type of energy has extreme negative pressure, leading to even more accelerated expansion of the universe compared to dark energy. The consequences of phantom energy can be quite dramatic, potentially resulting in a hypothetical scenario known as the "Big Rip," where the universe would be torn apart as the expansion accelerates infinitely.

$\bullet$ For the case when $\omega_{i}=-1$, it corresponds to a special kind of energy called the "cosmological constant" or "vacuum energy". A cosmological constant has a constant energy density throughout space and a pressure equal in magnitude but opposite in sign to its energy density. This energy component plays a crucial role in driving the accelerated expansion of the universe, as confirmed by observational evidence in the form of cosmic microwave background radiation and Type Ia supernovae data.

$\bullet$ Lastly, when $\omega_{i}<-1$, it introduces a concept known as "phantom dark energy". In this case, $a \varpropto\rho_{eff}^{\frac{2}{n}}$ and $a \varpropto t^n$, and $n<0$.  Phantom dark energy signifies a hypothetical energy form with an equation of state parameter less than -1, resulting in even more extreme negative pressure than phantom energy discussed earlier. This extreme exotic energy type could lead to bizarre scenarios such as the "Big Rip," where the universe would be torn apart in a finite time due to the repulsive nature of its gravitational effects.

On the other hand, in an expanding, homogeneous, and isotropic universe, the Hubble law states that the relative velocity of observer $B$ with respect to $A$ is given by 
\begin{eqnarray}
\bold{v}_{AB} =H(t) \bold{r}_{AB}\label{Hubble01},
\end {eqnarray}
where $H(t)$ is the Hubble parameter, and $\bold{r}_{BA}$ is the vector pointing from $A$ to $B$ \citep{mukhanov2005physical}. This expansion can be visualized using the analogy of an expanding sphere's surface, where the distance between points $A$ and $B$ increases as the radius $a(t)$ of the sphere grows, Fig. 1. This relationship is expressed as
\begin {eqnarray}
r_{AB}=a(t)\theta_{AB},
\end{eqnarray}
leading to a relative velocity
\begin {eqnarray}
v_{AB}=\dot{r}_{AB}=\dot{a}(t)\theta_{AB}= \frac{\dot{a}}{a}r_{AB},
\end{eqnarray}
where the dot denotes a derivative with respect to time. Thus, the Hubble law is represented by $H(t) \equiv \frac{\dot{a}}{a}$. In fact, the scale factor $a(t)$ describes the distance between observers as a function of time. 

\begin{tabular*}{2.5 cm}{cc}
\hspace{2.5 cm} \includegraphics[scale=.35]{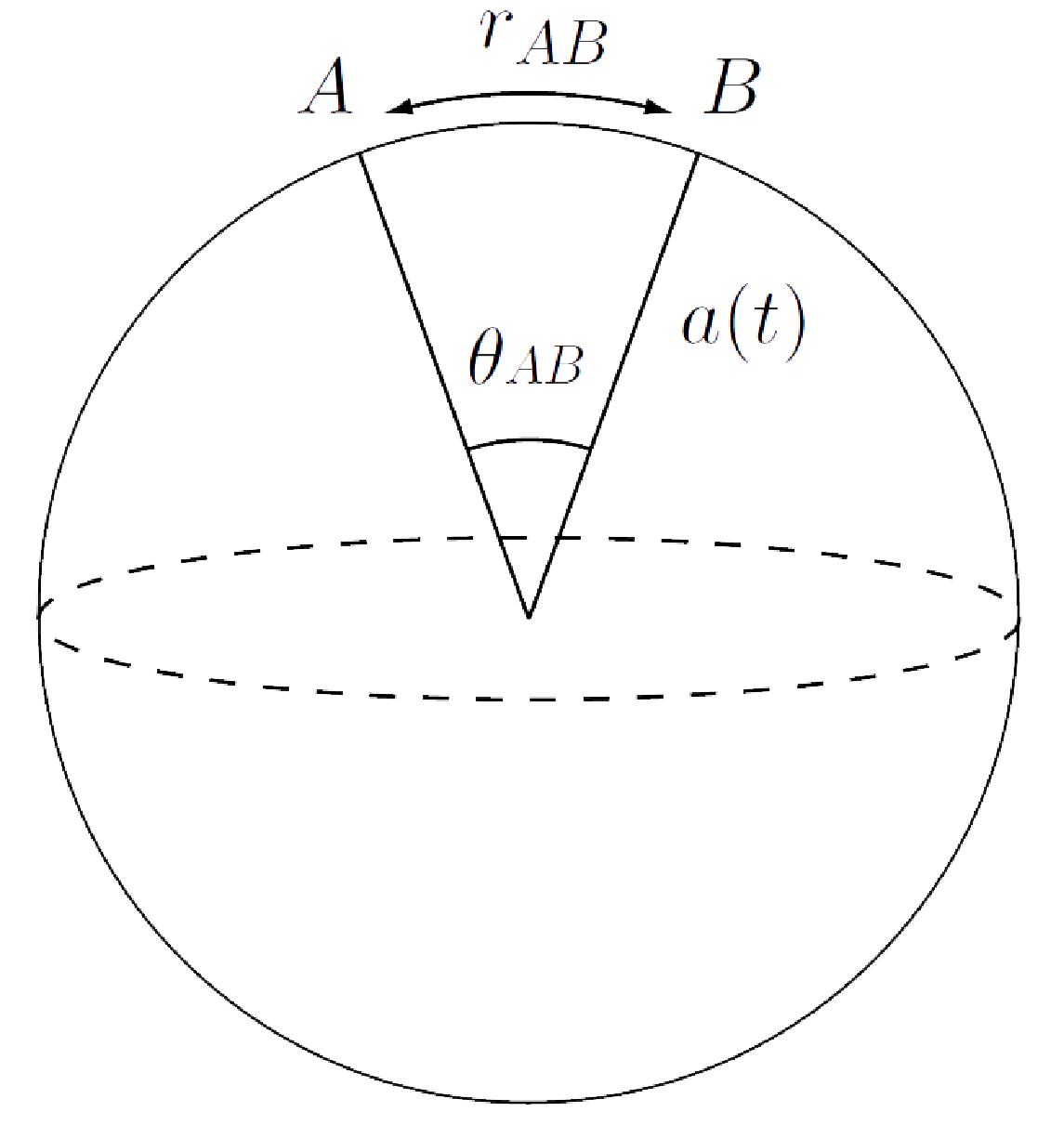}\\
\hspace{2.5 cm} Fig.1: Two-dimensional surface of an expanding sphere \\
\end{tabular*}\\

It means that if two observers travel relative to each other at the speed of light, $c$, the distance between them is given by $a(t)=ct$. This implies that only for $-1<\omega_i<-\frac{1}{3}$ where the values of $n$ are greater than $1$, the scale factor expands more rapidly than observers moving at the speed of light. This scenario occurred during the inflationary epoch, where space-time expanded faster than the speed of light. In contrast, in other cases, the rate of expansion of space-time remains below the speed of light.

\section{Weyl Conformal Geometry}\label{subsec:Weyl}

Under Weyl transformations $g_{{\mu\nu}}\rightarrow \Omega ^{2}(x)g_{{\mu\nu}}$ the conformal gravity is invariant under conformal transformations in the Riemannian geometry sense. Where $g_{\mu\nu}$ is the metric tensor and $\Omega (x)$ is a function on space-time. In this way, where $R^{\mu\nu\rho\lambda}R_{\mu\nu\rho\lambda}-4R^{\mu\nu}R_{\mu\nu}+R^{2}=G$ is a Gauss-Bonnet term, which has no quota in the equation of motion and vanishes during integrating, the contribution of $R^{\mu\nu\rho\lambda}R_{\mu\nu\rho\lambda}$ in the action,
can be written in terms of $R^2$ and $R_{\mu\nu}R^{\mu\nu}$. Thus where $C_{\mu\nu\rho\lambda}$ is the Weyl tensor:
\begin{eqnarray}\label{Weyl tensor}
C_{\mu\nu\rho\lambda}=R_{\mu\nu\lambda\rho}
-\frac{1}{2}(g_{\mu\lambda}R_{\nu\rho}-g_{\mu\rho}R_{\nu\lambda}-g_{\nu\lambda}R_{\mu\rho}+g_{\nu\rho}R_{\mu\lambda})
+\frac{R}{6}(g_{\mu\lambda}g_{\nu\rho}-g_{\mu\rho}g_{\nu\lambda})\cdot
\end{eqnarray} 
with replacing $R\rightarrow C_{\mu\nu\rho\lambda}C^{\mu\nu\rho\lambda}(R)+\Xi(\phi,\psi)$ in Einstein-Hilbert action, action is given by,

\begin{eqnarray}\label{ac2}
\mathcal{S}=\int{d^4x\sqrt{-g}\left(-\frac{R_{\mu\nu}R^{\mu\nu}-\frac{1}{3}R^{2}}{k^2}+\frac{\Xi(\phi,\psi)}{\kappa_s^2}+\mathcal{L}_{m,r}\right)}\cdot
\end{eqnarray}
The first term of the above action is equivalent to
\begin{eqnarray}
R_{\mu\nu}R^{\mu\nu}-\frac{1}{3}R^{2}=-12\left(H^2+\dot{H}\right)H^2=-\frac{1}{2}G\cdot
\end{eqnarray}
So, the function $f(R,G,T)$ is equivalent to $G$, and Eqs. (\ref{rho_R}) to (\ref{p_G}), and (\ref{rho-DE}) can be rewritten as
\begin{eqnarray}
\rho_{R-Weyl}&=&12H^2(\dot{H}+H^2) \label{rho_R_W}\\
p_{R-Weyl}&=&-12H^2(\dot{H}+H^2) \label{p_R_W}\\
\rho_{G-Weyl}&=&-12H^2(\dot{H}+H^2) \label{rho_G_W}\\
p_{G-Weyl}&=&12H^2(\dot{H}+H^2) \label{p_G_W}\\
k^2\rho_{DE-Weyl} &=& 3H^2+\frac{1}{m^2_p}\rho_{\Xi}\cdot\label{rho_DE_W}
\end{eqnarray}
Unfortunately, these equations show us that equations $\mathcal{W}_{00}$ and $\mathcal{W}_{ii}$ of my previous works, \citep{ghanaatian2014bouncing} and \citep{ghanaatian2018bouncing} are equal to zero. Consequently, their related results are wrong. They were my worst miscalculations, and I hope not to repeat any errors like them in my future work. 

To comply with the conservation of energy law, we need only verify the condition (\ref{conserve-condition}), where both sides of the equation equal zero, confirming that this model adheres to the law of energy conservation.

\section{A Single Inflaton and crossing Over the Phantom Divider}
During the inflationary epoch, if we consider a single inflaton scalar field, it rolls at a much slower rate than the expansion of the Universe. The inflaton potential energy is like a high viscose fluid, and the inflaton is like a falling ball in it.  So, the friction causes the inflaton to roll slowly along the potential with a constant velocity. In addition, to explain what has happened, it seems, we do not need the phantom scalar field to exist. As will be shown in the next section, this assumption does not satisfy $\omega_{eff}$ crossing over $-1$. It means that one can assume $V(\phi,\psi)=V(\psi)$, the accelerated term is negligible, $\ddot{\psi}\simeq 0$, and the inflaton potential energy is more than its kinetic energy, $V(\psi)\gg \dot{\psi}^2/2$. Therefore, 
\begin{eqnarray}
3H\dot{\psi}+V_{,\psi}\simeq 0, \label{EOM-psi-inflaton}
\end{eqnarray}
where $V_{,\psi}$ indicate differentiation of $V(\psi)$ with respect to the inflaton, $\psi$. 

According to inflationary models, the vacuum state of the universe was different from what it is today. At this time, the universe was dominated by vacuum energy. The early vacuum state had a much higher energy density, generating a repulsive force that caused a rapid expansion of space. This expansion explains many properties of the current universe that are difficult to account for without an inflationary epoch.

On the other hand, Ricci curvature is a mathematical concept related to space geometry. While Ricci curvature is crucial for understanding space-time dynamics in the presence of matter and energy, it is not directly tied to the inflationary epoch itself, so $\rho_{vac}\gg \rho_{R}$ and $\rho_{vac}\gg \rho_{G}$.

During inflation, the energy density of the inflaton field remains approximately constant, although it tends to dilute over time. Eventually, the energy density of the inflaton field is transferred to other degrees of freedom during reheating.

In summary, the energy density of the inflaton field and the vacuum energy density are closely related during the inflationary epoch. The inflaton field dominates the energy density, driving the universe's rapid expansion. We have $p_{vac}\gg p_{R}$ for similar reasons. As the inflaton field decays, its energy is transferred to other fields like the phantom scalar fields, marking the end of the inflationary period and the start of the radiation-dominated era of the universe. So, the first modified equation of Friedmann, eq.(\ref{f1-E}), using eqs. (\ref{rho_Xi}) and (\ref{rho-DE}), is rewritten approximately by
\begin{eqnarray}
H^2\simeq \frac{V(\psi)}{3m_p^2},
\end{eqnarray}
therefore the consistency conditions are
\begin{eqnarray}
\left\vert m_{p}\frac{V_{,\psi}}{V(\psi)}\right\vert \ll \sqrt{6},\qquad\text{and}\qquad \left\vert m_{p}^{2}\frac{V_{,\psi\psi}}{V(\psi)}\right\vert \ll 3. 
\end{eqnarray}
These two consistency conditions demand an extremely flat potential for inflaton, which also results in a large amount of inflation or e-folding (N), which is defined by
\begin{eqnarray}
N\doteq\int_{t_{i}}^{t_{f}}{H(t)dt}=\int^{\psi_{i}}_{\psi_{f}}{\frac{1}{m^2_p}\frac{V(\psi)}{V_{,\psi}}d\psi},
\end{eqnarray}
where $t_{i}$ and $t_{f}$ are the time at the beginning and the end of the inflation period. The Friedmann equations can be written in terms of the Hubble parameter as
\begin{eqnarray}
H^2&=&\frac{1}{3m^2_p}\left(\frac{\dot{\psi}^2}{2}+V(\psi)\right),\label{f1-inflation}\\
\dot{H}&=&\frac{-1}{2m^2_p}\dot{\psi}^2.\label{f2t-inflation}
\end{eqnarray}
Using eqs. (\ref{f1-inflation}) and (\ref{f2t-inflation}), one can obtain
\begin{eqnarray}
V(\psi)=3m_p^2H^2\left(1-\frac{1}{3}\epsilon\right),
\end{eqnarray}
where 
\begin{eqnarray}\label{epsilon}
\epsilon\doteq\epsilon(\psi)= 2m^2_p\left(\frac{H_{,\psi}}{H(\psi)}\right)^2=-\frac{\dot{H}}{H^2}=\frac{1}{2m^2_p}\left(\frac{\dot{\psi}}{H(\psi)}\right)^2,
\end{eqnarray}
is a slow roll parameter, and
\begin{eqnarray}
H_{,\psi}&\doteq&\frac{\partial H}{\partial\psi}=\frac{-1}{2m^2_p}\dot{\psi}.\label{f2psi-inflation}
\end{eqnarray}
The first and second derivatives of the potential with respect to inflaton scalar field are
\begin{eqnarray}
V_{,\psi}&=&3\sqrt{2}m_p H^2\sqrt{\epsilon}\left(1+\frac{1}{3}\delta_1\right),\label{Psi_{,psi}}\\
V_{,\psi\psi}&=&3H^2\left(\epsilon-\delta_1-\frac{1}{3}\delta_2\right),\label{Psi_{,psipsi}}
\end{eqnarray}
here,
\begin{eqnarray}
\delta_{1}&=&-2m_p^2\frac{H_{,\psi\psi}}{H}=\frac{\ddot{\psi}}{H\dot{\psi}}\label{delta1}\\
\delta_{2}&=&\frac{4m_p^2}{H^2}\left(H_{,\psi}H_{,\psi\psi\psi}+H_{,\psi\psi}^{2}\right)=\frac{\dddot{\psi}}{H^2\dot{\psi}}\label{delta2}\\
\delta_n&=&\frac{1}{H^n \dot{\psi}}\frac{d^{n+1}\psi}{d^{n+1}t}\cdot
\end{eqnarray}
By finding the rate of change of the Hubble parameter over time, $\dot{H}=\frac{\ddot{a}}{a}-\left(\frac{\dot{a}}{a}\right)^2$, and using (\ref{epsilon}), we have   
\begin{eqnarray}
\frac{\ddot{a}}{a}=H^2+\dot{H}=H^2\left(1+\frac{\dot{H}}{H^2}\right)=H^2(1-\epsilon).\label{a2t}
\end{eqnarray}
If $\epsilon<1$, then it is necessary that $\ddot{a}>0$ for inflation to occur. Inflation refers to the expansion of the universe during its early stages. It is necessary to have a significant amount of inflation to magnify quantum fluctuations to the extent that they form the large-scale structure that we see today. To achieve this, we need $\epsilon \ll 1$ or $\epsilon\rightarrow 0$ and also $\delta_n <1$. From equations (\ref{Psi_{,psi}}) and (\ref{Psi_{,psipsi}}), we can infer that the slow roll parameters must satisfy the consistency conditions, which impose limits on $V_{,\psi}\ll 1$, and $V_{,\psi\psi}\ll 1$.

Using the equations (\ref{f1-E}), (\ref{f2-E}), (\ref{rho_Xi}), (\ref{p_Xi}), and (\ref{epsilon}) in the inflation method, one can rewrite energy density, pressure, and the EoS parameter as
\begin{eqnarray}
\rho_{\psi}&=&\,\,\,\,\,3m^2_pH^2=\frac{\dot{\psi}^2}{2}+V(\psi),\label{rh0-inflation}\\
p_{\psi}&=&-3m^2_pH^2-2m^2_p\dot{H}=\frac{\dot{\psi}^2}{2}-V(\psi),\label{p-inflation}\\
\omega_{\psi}&=&\frac{p_{\psi}}{\rho_{\psi}}=-1-\frac{2}{3}\frac{\dot{H}}{H^2}=-1+\frac{2}{3}\epsilon.\label{omega-inflation}
\end{eqnarray}
If $\epsilon$ approaches $0$, then $\omega_{\psi}$ tends towards $-1$, which is the phantom divider from the quintessence. If $\epsilon>0$, then $\omega_{\psi}$ must be greater than $-1$, indicating the quintessence, whereas, for $\epsilon<0$ and $\omega_{\psi}<-1$, we face phantom. On the other hand, the equation (\ref{epsilon}) shows that $\epsilon$ is never gotten negative, because the inflaton scalar field is a real field and $\dot{\psi}^{2}$ is always positive. Therefore, although the inflation model satisfies the bouncing conditions for the scale factor and the Hubble parameter, the equation of state of this model cannot cross over $-1$.

\section{Numerical solution for a successful bouncing}\label{sec:NSSB}
In this section, I will numerically analyze the early-time behavior of cosmological parameters, such as $\omega$. Furthermore, I am attempting to identify the factors contributing to a successful bounce using simple sample models. To understand the dynamics of the universe, we need to study the variations of the Hubble parameter or scale factor with respect to cosmic time, denoted as $t$. For a successful bounce, we should observe the following behaviors:

- A contraction for $t<0$, meaning that the scale factor $a(t)$ should decrease ($\dot{a}<0$).

- A bounce at $t=0$, where $\dot{a} = 0$. Around this point, $\ddot{a} > 0 $ and the Hubble parameter $H$ changes from negative to positive, reaching zero at the bouncing point.

- An expansion for $t>0$, meaning that the scale factor $a(t)$ should increase or $\dot{a}>0$.

\subsection{Inflation model vs. Quinton model}
In the simplest case, by using $f(R,G,T)=R$, and $\mathcal{L}_{m,r}=0$, the eq. (\ref{ac1}) can be rewritten as
\begin{eqnarray}\label{ac-Num-01}
\mathcal{S}=\frac{1}{2k^2}\int{d^4x\sqrt{-g}\left(R+\frac{2}{m^2_p}\Xi(\phi,\psi)\right)},
\end{eqnarray}
In numerical methods for solving equations, adding each parameter increases the number of effective variables. In this context, there are three independent variables: the scale factor, $a(t)$, the phantom scalar field, $\phi(t)$, and the inflaton scalar field, $\psi(t)$. So, by combining the second Friedmann equation (\ref{f22}), with two wave motion equations (\ref{EOM-phi}), and (\ref{EOM-psi}), numerical values for the variables are obtained. Our model successfully demonstrates a bounce, as shown in Figures 2 and 3. Figure 2 features a single scalar inflaton field, while Figure 3 incorporates both an inflaton and a phantom field in a quintom model.

\begin{tabular*}{2.5 cm}{cc}
\includegraphics[scale=.35]{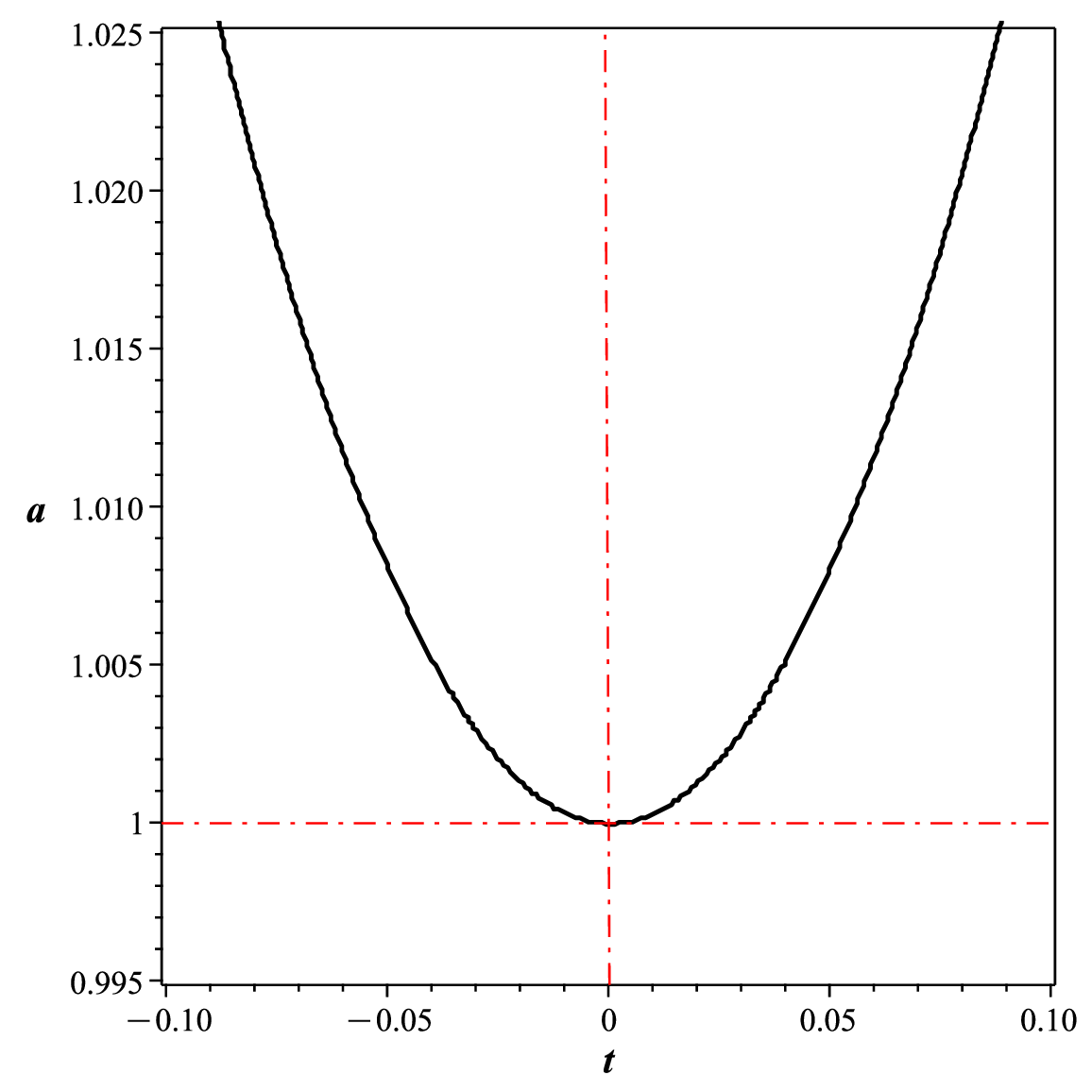}\hspace{1 cm}\includegraphics[scale=.35]{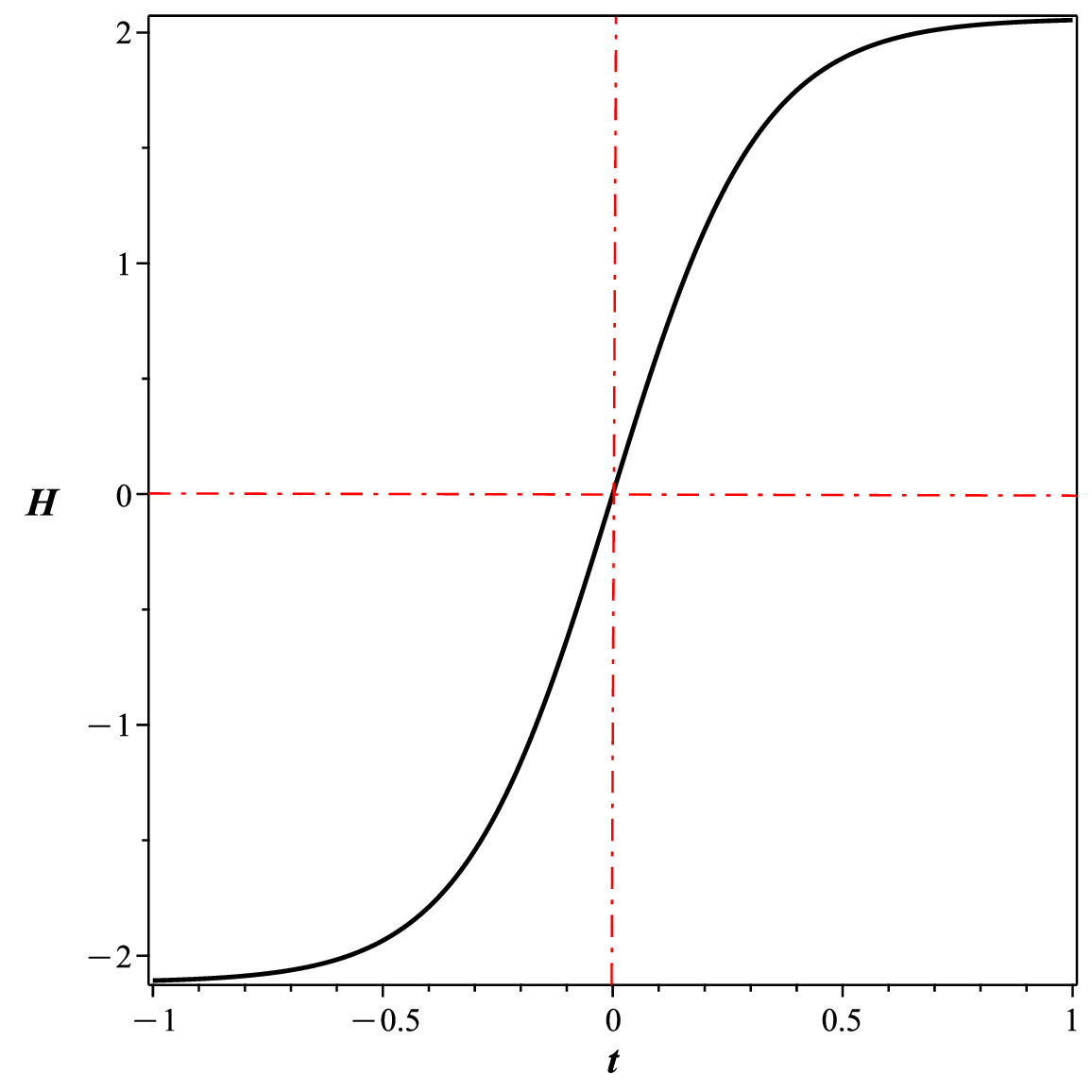}\hspace{1 cm}\\
\hspace{1 cm} Fig.2: The graph of the scale factor, $a$, and Hubble parameter, $H$, \\
as the functions of time, for $V= V_{0}e^{\beta\psi}$, where $G=1$, $c=1$, $m_{p}=\frac{1}{\sqrt{8\pi}}$, $k=\sqrt{8\pi}$, \\
$V_0=0.5$, $\beta=-\frac{\sqrt{2}}{3}$, and the initial values are, $\psi(0) =-0.05$, $\dot{\psi}(0) = 0.1$, and $\dot{a}(0)=0$.\\
\end{tabular*}\\

\begin{tabular*}{2.5 cm}{cc}
\includegraphics[scale=.35]{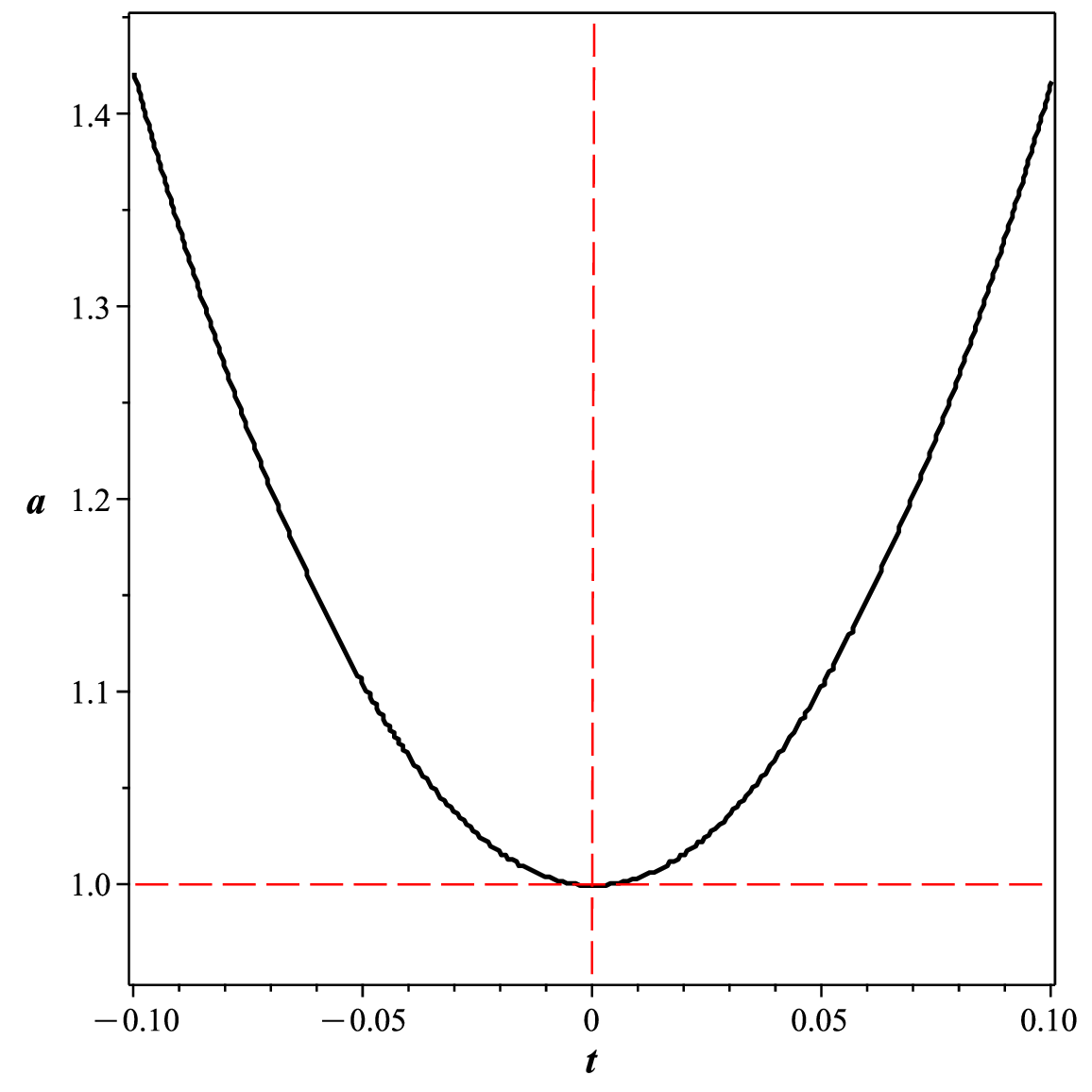}\hspace{1 cm}\includegraphics[scale=.35]{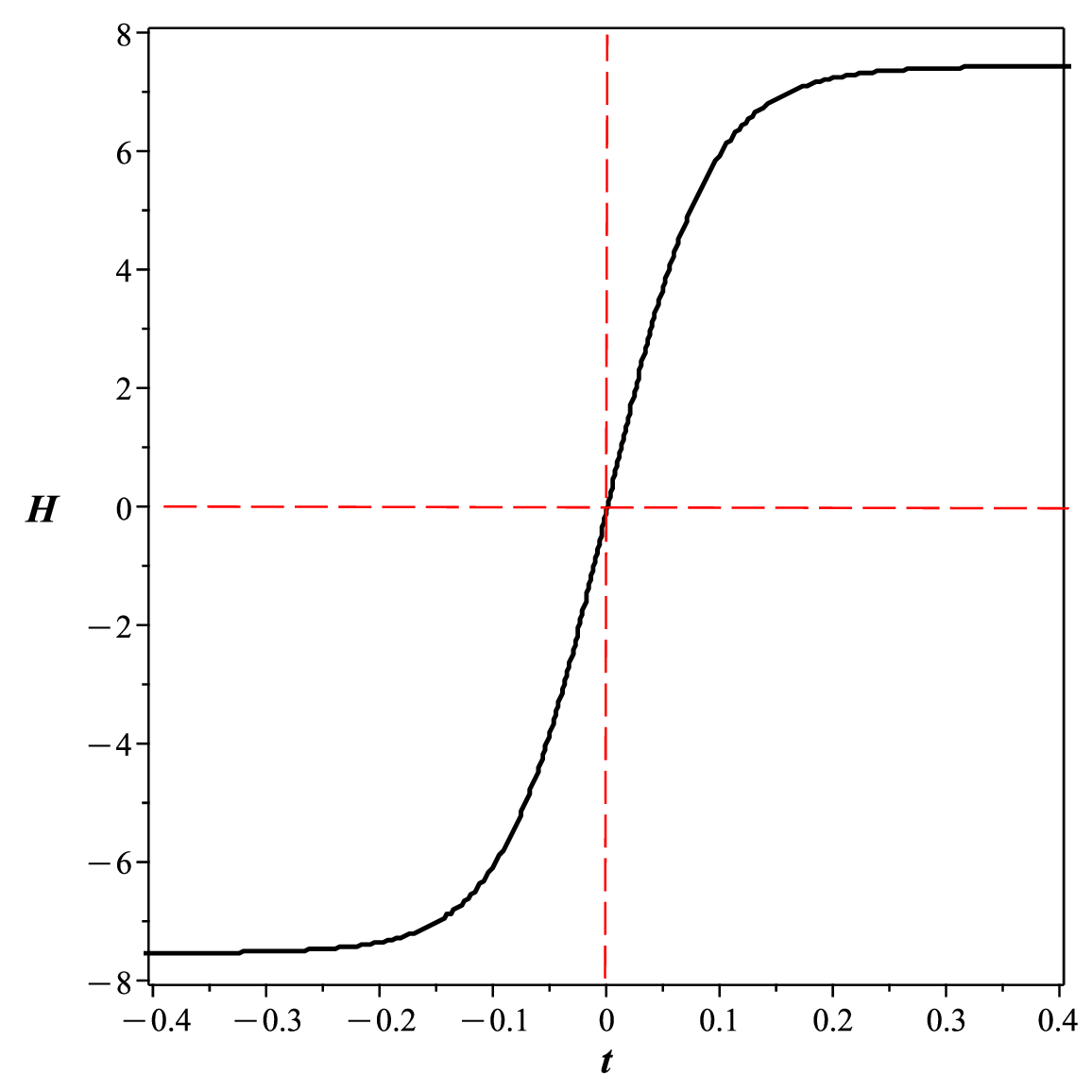}\hspace{1 cm}\\
\hspace{1 cm} Fig.3: The graph of the scale factor, $a$, and Hubble parameter, $H$, \\
as the functions of time, for $V= V_{0}e^{\alpha\phi+\beta\psi}$, where $G=1$, $c=1$, $m_{p}=\frac{1}{\sqrt{8\pi}}$, $k=\sqrt{8\pi}$, \\
$V_0=0.25$, $\alpha=-\frac{\sqrt{6}}{3}$, $\beta=-1$, and the initial values are, $\phi(0) =-0.05$, $\dot{\phi}(0) = 0.1$,\\
$\psi(0) =0.05$, $\dot{\psi}(0) = -0.1$, and $\dot{a}(0)=0$.\\
\end{tabular*}\\

As we mentioned, in models known as "quintom," $\omega_{eff}$ varies over time, initially having a value below $-1$ to match an accelerating expansion of the universe and then later transitioning to a value above $-1$ to correspond to a potential future shrinking phase. As a result, quintom provides a model in which the universe expands from a Big Bang, experiences an accelerating rate of expansion, and may later undergo a collapse into a Big Crunch, allowing for the possibility of a Bounce cycle.

In Figure 4, we have a single inflaton scalar field. The EoS parameter does not cross over $-1$. The $\omega_{eff}$ closes to $-1$ from left, $\dot{H}$ are positive and the scale factor $a(t)$ is not zero at the bouncing point. It sees its peak at the bouncing point and then it decreases from right asymptotically until $-1$. 

\begin{tabular*}{2.5 cm}{cc}
\includegraphics[scale=.35]{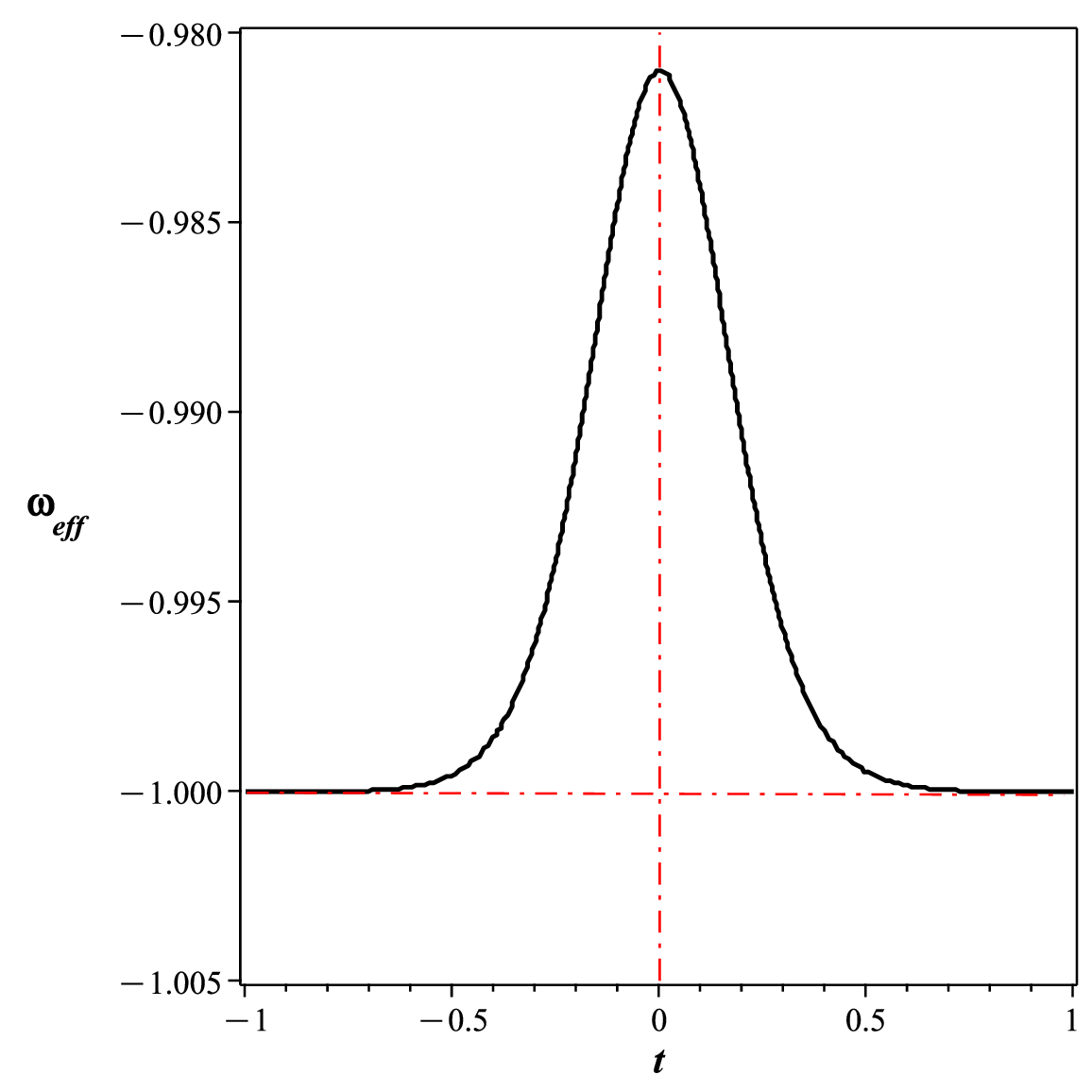}\hspace{1 cm}\includegraphics[scale=.35]{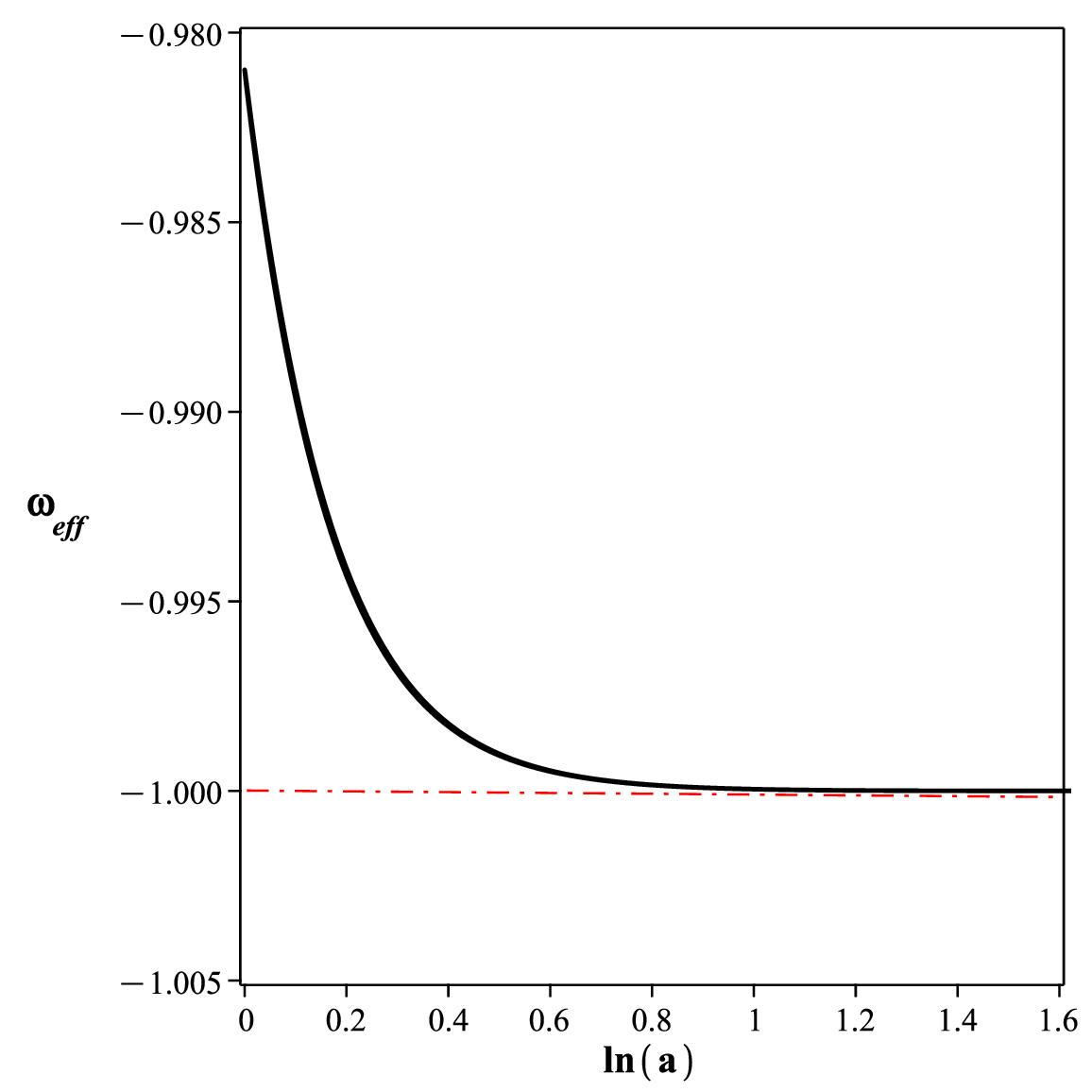}\hspace{1 cm}\\
\hspace{1 cm} Fig.4: The graph of the equation of state, $\omega$, respect to the time and $\ln(a)$\\
for $V= V_{0}e^{-\beta\psi}$, where $G=1$, $c=1$, $m_{p}=\frac{1}{\sqrt{8\pi}}$, $k=\sqrt{8\pi}$, \\
$V_0=0.5$, $\beta=\frac{\sqrt{2}}{3}$, and the initial values are, $\psi(0) =-0.05$, $\dot{\psi}(0) = 0.1$, and $\dot{a}(0)=0$.\\
\end{tabular*}\\

While, in Figure 5, we consider an inflaton and a phantom scalar filed together in a quintom model. The EoS parameter crosses over $-1$ from $\omega_{eff}<-1$ to $\omega_{eff}>-1$. The $\omega_{eff}$ closes to $-1$ from left, $\dot{H}$ are positive and the scale factor $a(t)$ is not zero at the bouncing point. Therefore, we've avoided the singularity that was faced in the usual Einstein cosmology.

\begin{tabular*}{2.5 cm}{cc}
\includegraphics[scale=.35]{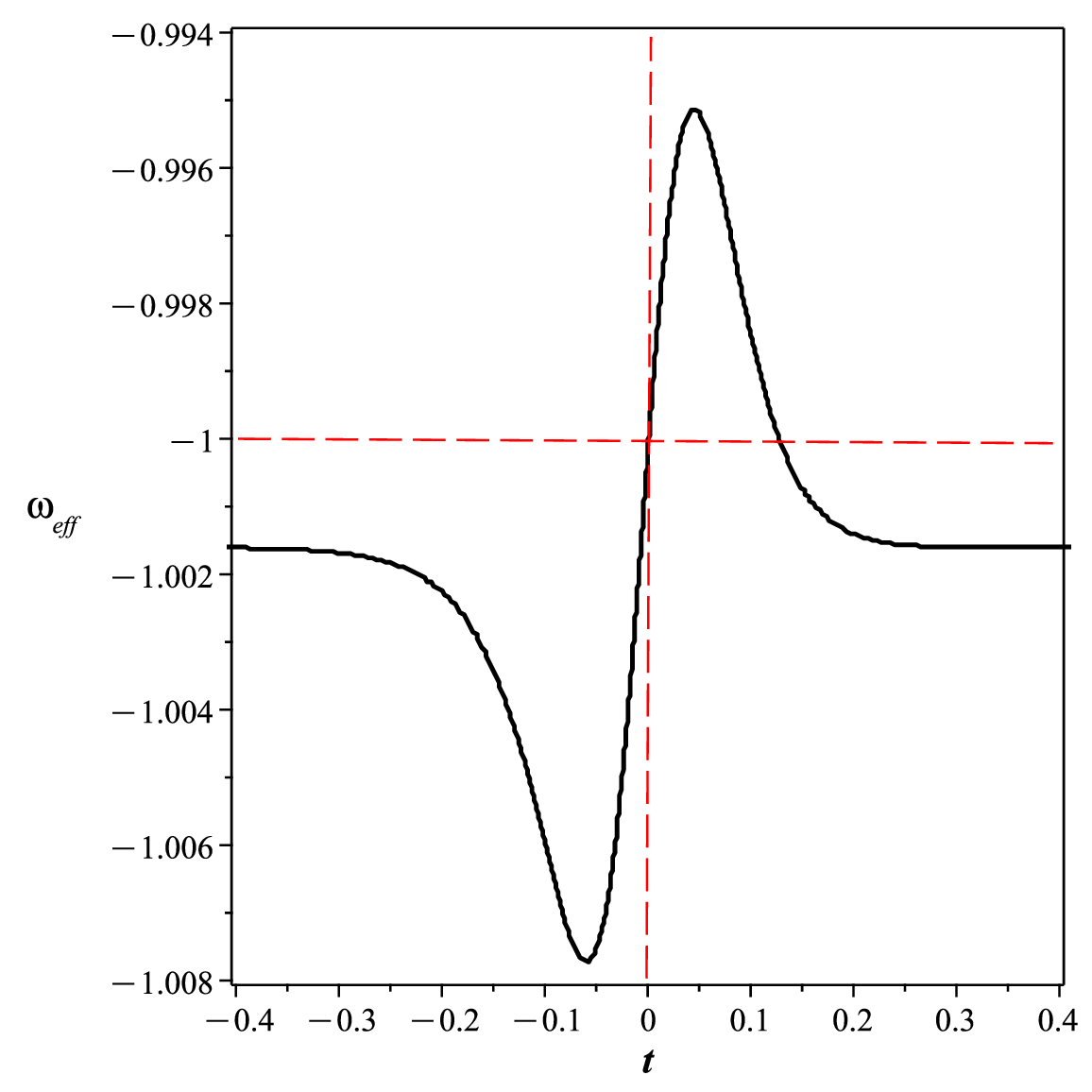}\hspace{1 cm}\includegraphics[scale=.35]{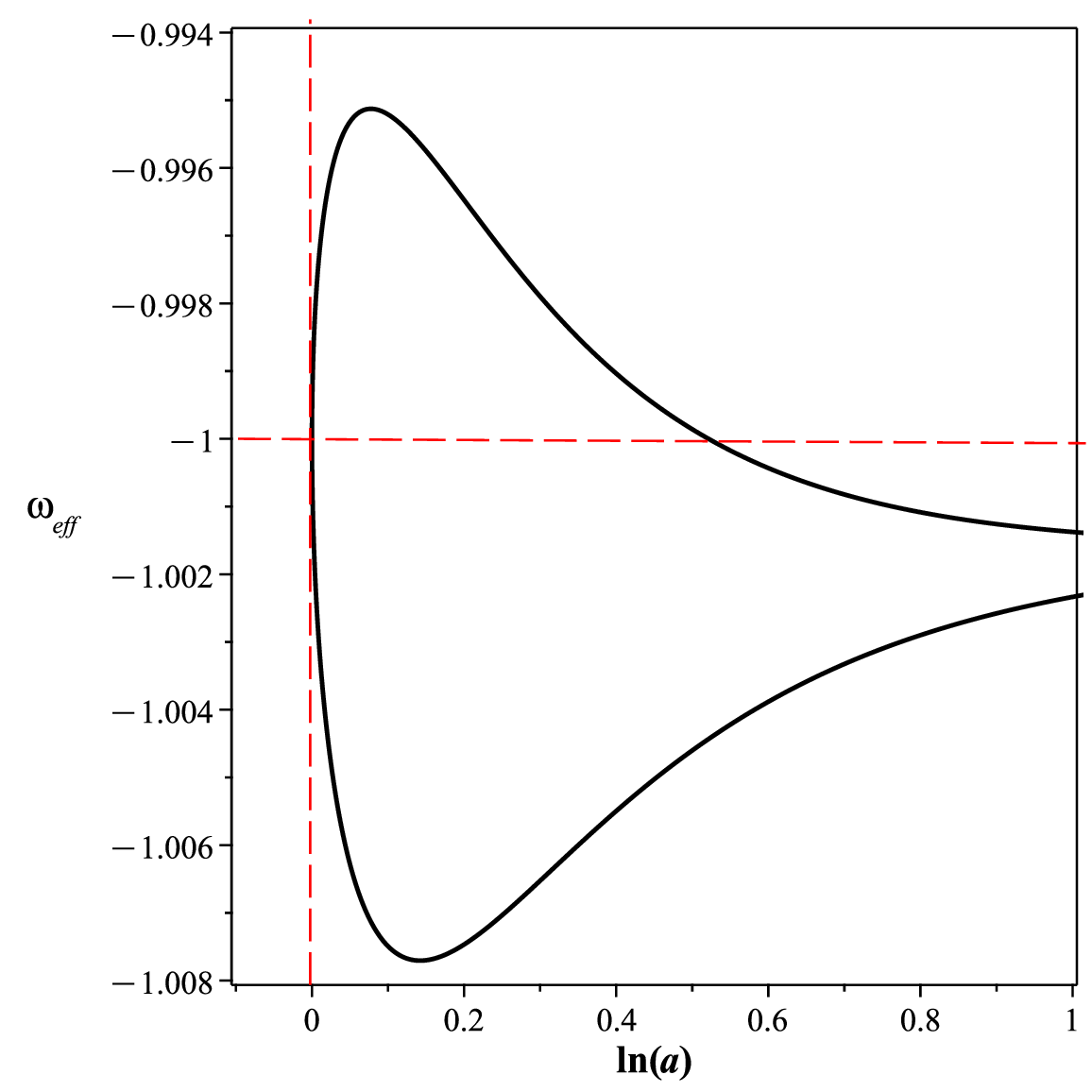}\hspace{1 cm}\\
\hspace{1 cm} Fig.5: The graph of the equation of state, $\omega$, respect to the time and $\ln(a)$\\
for $V= V_{0}e^{-\alpha\phi-\beta\psi}$, where $G=1$, $c=1$, $m_{p}=\frac{1}{\sqrt{8\pi}}$, $k=\sqrt{8\pi}$, \\
$V_0=0.25$, $\alpha=\frac{\sqrt{6}}{3}$, $\beta=1$, and the initial values are, $\phi(0) =-0.05$, $\dot{\phi}(0) = 0.1$,\\
$\psi(0) =0.05$, $\dot{\psi}(0) = -0.1$, and $\dot{a}(0)=0$.\\
\end{tabular*}\\

\subsection{What will happen if $\mathcal{L}_{m,r}$ is not negligible?}

To avoid complicating the equations too much, we assume $f(R, G, T)=R$, and $\mathcal{L}_{m,r}\neq 0$. So, the eq. (\ref{ac1}) can be rewritten as
\begin{eqnarray}\label{ac-Num-01}
\mathcal{S}=\int{d^4x\sqrt{-g}\left(\frac{1}{2k^2}R+\frac{1}{\kappa^2_s}\Xi(\phi,\psi)+\mathcal{L}_{m,r}\right)},
\end{eqnarray}
This introduces two additional time-dependent variables to our previous variables: $\rho_{m,r}$, and $p_{m,r}$. Consequently, we require two additional equations for numerical solutions, focusing specifically on the quintom model. The bouncing conditions are demonstrated in Figure 6.  

\begin{tabular*}{2.5 cm}{cc}
\includegraphics[scale=.35]{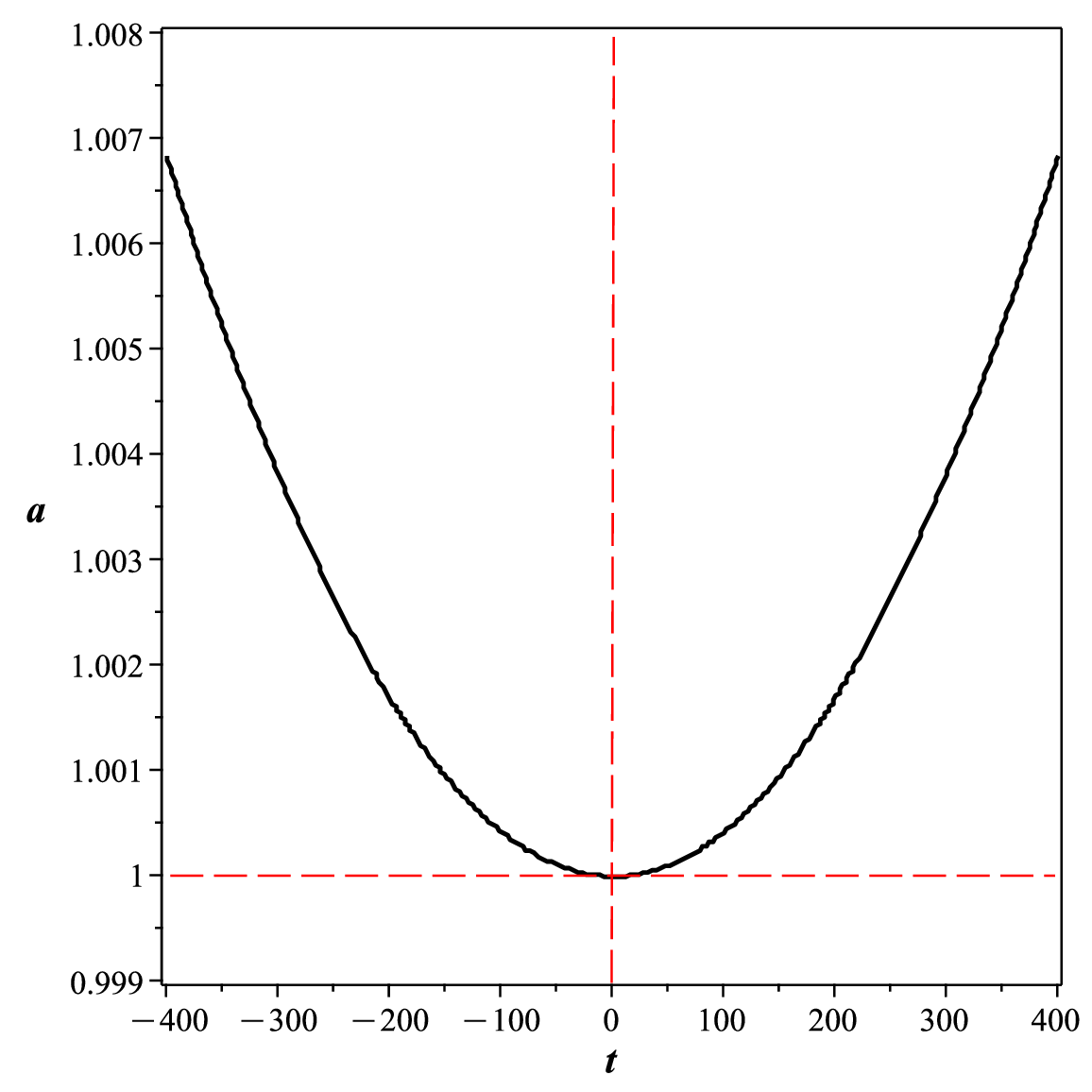}\hspace{1 cm}\includegraphics[scale=.35]{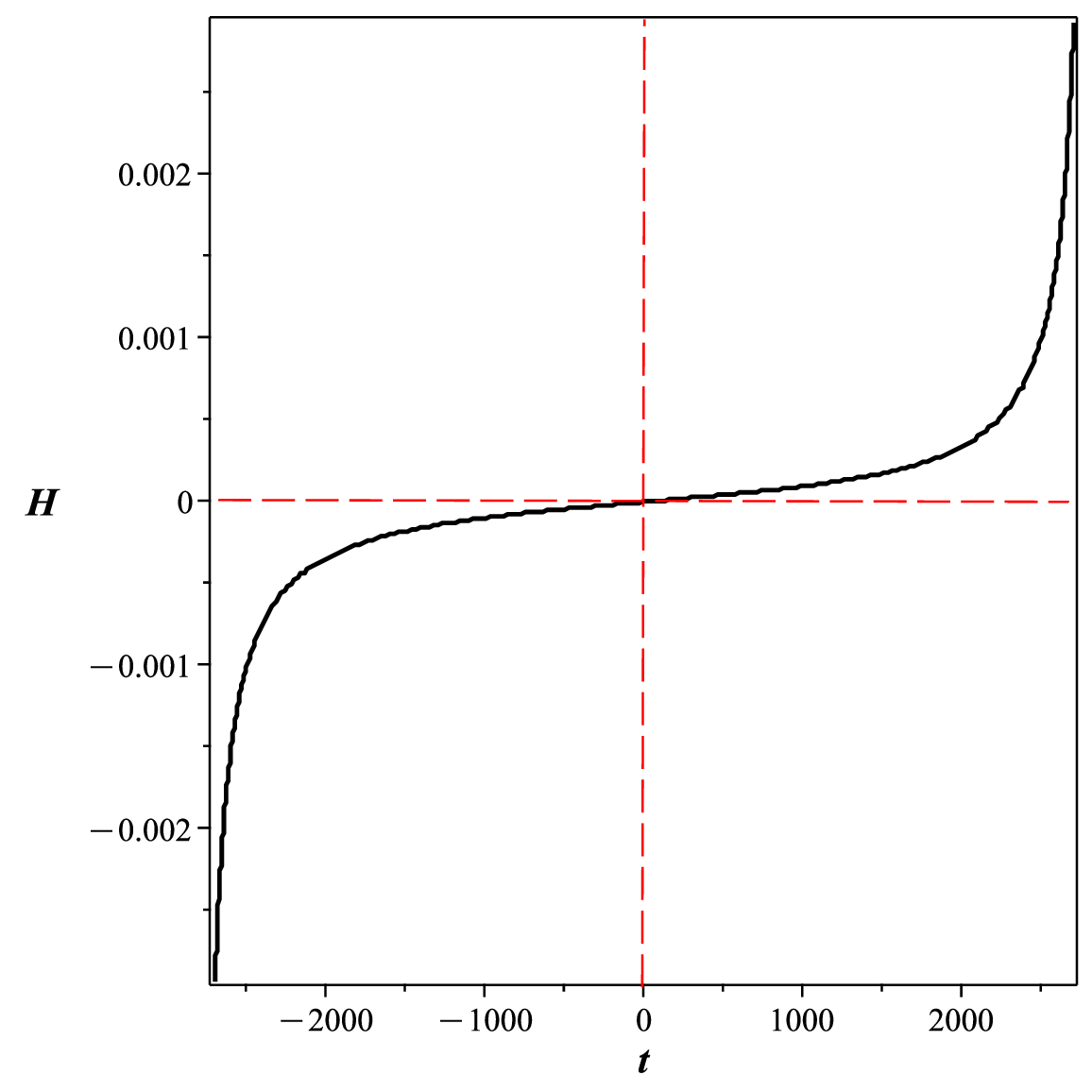}\hspace{1 cm}\\
\hspace{1 cm} Fig.6: The graph of the scale factor, $a$, and Hubble parameter, $H$, \\
as the functions of time, for $V= V_{0}e^{-\alpha\phi-\beta\psi}$, where $G=1$, $c=1$, $\kappa_{s}=1$, $k=\sqrt{8\pi}$, \\
$V_0=2.5\times 10^{121}$, $\alpha=2$, $\beta=1$, and the initial values are, $\phi(0) =50$, $\dot{\phi}(0) = 0.001$,\\
$\psi(0) =200$, $\dot{\psi}(0) = -0.001$, $\rho_{m,r}(0)=-1\times 10^{-9}$, $p_{m,r}(0)=1.02 \times 10^{-9}$, and $\dot{a}(0)=0$.\\
\end{tabular*}\\

In Figure 7, the EoS parameter crosses over $-1$ twice. Once, before bouncing point, and other after that. The $\omega_{eff}$ closes to $-1$ from left, $\dot{H}$ are positive and the scale factor $a(t)$ is not zero at the bouncing point. Again, it moves away from $-1$ from the right after bouncing point. 

\begin{tabular*}{2.5 cm}{cc}
\includegraphics[scale=.35]{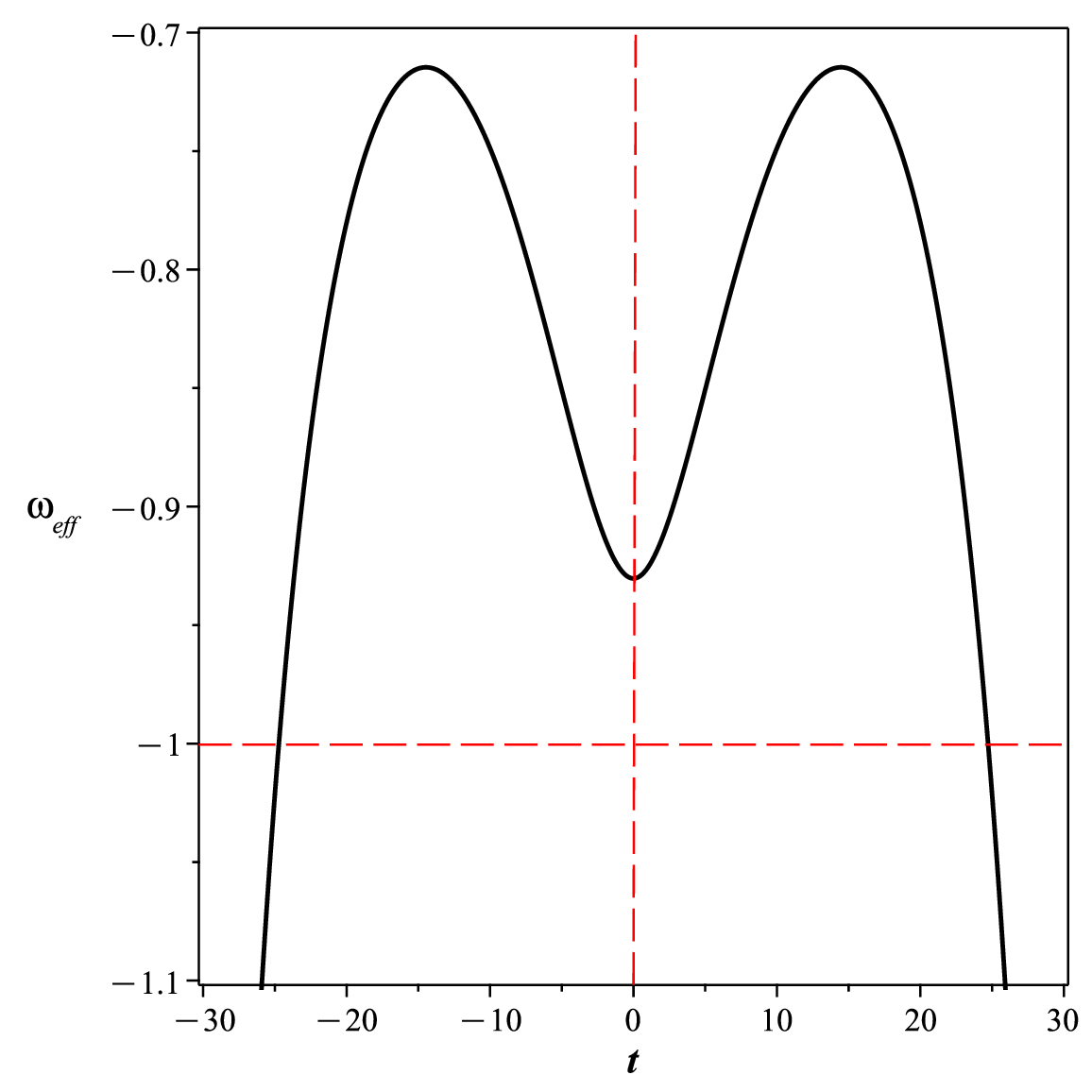}\hspace{1 cm}\includegraphics[scale=.35]{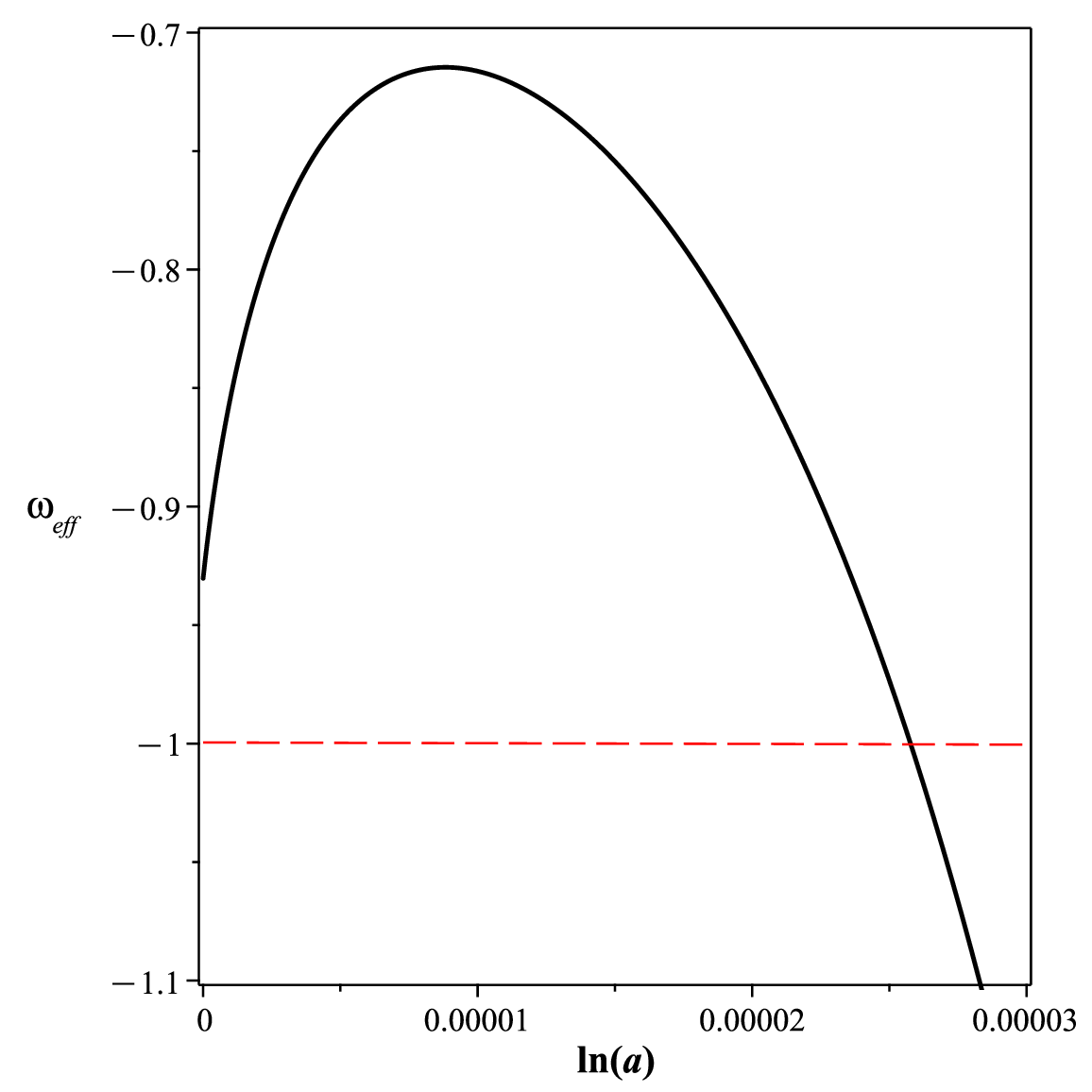}\hspace{1 cm}\\
\hspace{1 cm} Fig.7: The graph of the equation of state, $\omega$, respect to the time and $\ln(a)$\\
for $V= V_{0}e^{-\alpha\phi-\beta\psi}$, where $G=1$, $c=1$, $\kappa_{s}=1$, $k=\sqrt{8\pi}$, \\
$V_0=2.5\times 10^{121}$, $\alpha=2$, $\beta=1$, and the initial values are, $\phi(0) =50$, $\dot{\phi}(0) = 0.001$,\\
$\psi(0) =200$, $\dot{\psi}(0) = -0.001$, $\rho_{m,r}(0)=-1\times 10^{-9}$, $p_{m,r}(0)=1.02 \times 10^{-9}$, and $\dot{a}(0)=0$.\\
\end{tabular*}\\

\section{Summary and Conclusion}\label{sec:SumCon}
In this article, I consider a $f(R, G, T)$ modified gravity model. This model is coupled with two distinct types of scalar fields. These fields are the phantom and the quintessence scalar fields if we consider the quintom model, and the phantom and the inflaton scalar fields if we want to study our model during the period of inflation. 

In section 2, the necessary conditions for a successful bounce in the flat FLRW universe were achieved. The wave equations of motion and the non-zero components of the energy-momentum tensor of the model were used to derive the modified Friedmann equations. This led to obtaining the energy density, and pressure of the dark energy. The analysis focused on the conditions in which to achieve a successful bounce,  $\omega_{eff}$ crosses over $-1$. Additionally, the study delves into the requirements for ensuring the conservation of energy law in general cases. 

In Section 3,  by referencing one of the collections of the Planck 2018 report, and estimating the energy density of total matter, baryonic matter, cold dark matter, radiation, and dark energy at the present time, the absence of radiation domination was confirmed.  In the absence of radiation, the relative of the scale factor with the effective energy density, equation of state, and cosmic time was achieved. Those relations were explored, by the various cases of the equation of states. The review of the Hubble law showed that only in the case of $-1<\omega_i<-\frac{1}{3}$, the scale factor expands more rapidly than observers moving at the speed of light. This scenario occurres during the inflationary epoch, when space-time expands faster than the speed of light. In contrast, in other cases, the rate of expansion of space-time remains below the speed of light.  

I emphasize that this model is a general model, encompassing other models.  In section 4, Weyl conformal geometry is the first example covered by this model. 

In section 5, examining the inflationary epoch for a single inflaton scalar field is the second example analyzed in my model. Under these circumstances,  when only the inflaton scalar field is considered, it was demonstrated that although the inflation model satisfies the bouncing conditions for the scale factor and the Hubble parameter, the EoS parameter cannot cross the phantom divider. 

Finally, in section 6, I try using two simple sample models. In section 6.1, as the simplest model, we conclude through numerical calculations that the inflation model, despite allowing for a bounce for the scale factor, cannot explain crossing over -1 for the state parameter, even in its simplest form. In contrast, the quintom model successfully meets the criteria for a successful bounce and clearly demonstrates the state parameter's cross over -1.
Therefore, I select the quintom model for further testing. By incorporating a non-minimal Lagrangian density of matter and radiation into our previous model, in section 6.2, we introduce two additional concepts. This model successfully explains both the bounce and the crossing over -1 for the state parameter. So, in section 6, the numerical solutions and graphs confirm the results obtained in sections 3 and 5.

Certainly, there are many potential models yet to be tested, and exploring these models in relation to observational data remains an open problem for future research.

\section{Appendix}\label{Appendix}
\subsection{Modified Stress-Energy Tensor}\label{subsec:Modified E-M-T}
The variation of the determinant of the metric, $\sqrt{-g}$, Ricci scalar, 
$R=g^{\mu\nu}R_{\mu\nu}$, Ricci tensor, $R_{\mu\nu}= 
\partial_{\alpha}\! \left({\Gamma{^{\alpha}_{\mu\nu}}}\right)-{\partial{_{\nu}}}\! \left({\Gamma{^{\alpha}_{\alpha\mu}}}\right)+{\Gamma{^{\beta}_{\mu\nu}}} {\Gamma{^{\alpha}_{\alpha\beta}}}-{\Gamma{^{\beta}_{\alpha\mu}}} {\Gamma{^{\alpha}_{\beta\nu}}}$, Gauss-Bonnet invariant, $G=R^{2}-4R_{\mu\nu}R^{\mu\nu}+R_{\mu\nu\rho\lambda}R^{\mu\nu\rho\lambda}$, and the trace of stress-energy tensor of the matter and radiation, $T=g^{\mu\nu}T_{\mu\nu}^{(m,r)}$, with respect to the inverse metric $g^{\mu\nu}$ is given by
\begin{eqnarray}
\delta\sqrt{-g}&=&-\frac{1}{2}\sqrt{-g}g_{\mu\nu}\delta g^{\mu\nu},\label{metric var}\\
\delta R = \frac{\partial\left(g^{\alpha\beta}R_{\alpha\beta}\right)}{\partial g^{\mu\nu}}\delta g^{\mu\nu}&=&R_{\mu\nu}\delta g^{\mu\nu}+g^{\alpha\beta}\frac{\partial R_{\alpha\beta}}{\partial g^{\mu\nu}}\delta g^{\mu\nu}\nonumber\\
&=&R_{\mu\nu}\delta g^{\mu\nu}+g_{\mu\nu}\square\delta g^{\mu\nu}-\nabla_\mu\nabla_\nu\delta g^{\mu\nu},\label{R var}\\ 
\delta R_{\mu\nu}&=& \nabla_\rho \delta \Gamma^{\rho}_{\nu\mu}-\nabla_\nu \delta \Gamma^{\rho}_{\rho\mu}\label{Rmunu var}\\
\delta G &=& 2R \delta R -4\delta (R_{\mu\nu}R^{\mu\nu})+\delta (R_{\mu\nu\rho\lambda}R^{\mu\nu\rho\lambda})\label{G var}\\
\delta T= \frac{\partial\left(g^{\alpha\beta}T_{\alpha\beta}^{(m,r)}\right)}{\partial g^{\mu\nu}}\delta g^{\mu\nu}&=&(T_{\mu\nu}^{(m,r)}+\Theta_{\mu\nu})\delta g^{\mu\nu}\label{T var}
\end{eqnarray}
where $\Theta_{\mu\nu}=g^{\alpha\beta}\frac{\partial T_{\alpha\beta}^{(m,r)}}{\partial g^{\mu\nu}}$, $\square=g^{\mu\nu}\nabla_\mu\nabla_\nu=\frac{1}{\sqrt{-g}}\partial_{\mu}\sqrt{-g}g^{\mu\nu}\partial_{\nu}$ is the d'Alembert operator, and $\delta \Gamma^{\lambda}_{\mu\nu}$ is the difference of two connections, it should transform as a tensor. Therefore, it can be written as
\begin{eqnarray}
\delta \Gamma^{\lambda}_{\mu\nu}=\frac{1}{2}g^{\lambda\alpha}\left(\nabla_{\mu}\delta g_{\alpha\nu}+\nabla_{\nu}\delta g_{\alpha\mu}-\nabla_{\alpha}\delta g_{\mu\nu}\right),
\end{eqnarray}
The variation of the action (\ref{ac1}) with respect to inverse metric  $g^{\mu\nu}$ is yielded by
\begin{eqnarray}
\delta S_{[g^{\mu\nu}]}&=&\frac{1}{2k^2} \int{\sqrt{-g}\left(f_R \delta R+f_G\delta G+f_T\delta T+\frac{f}{\sqrt{-g}}\delta\sqrt{-g}\right)d^4x}\nonumber\\
&+&\frac{1}{\kappa^2_s}\,\,\int{\sqrt{-g}\left(\frac{\partial \Xi(\phi,\psi)}{\partial g^{\mu\nu}}+\frac{\Xi(\phi,\psi)}{\sqrt{-g}}\frac{\partial\sqrt{-g}}{\partial g^{\mu\nu}}\right)\delta g^{\mu\nu}d^4x}\nonumber\\
&+&\,\,\,\,\,\,\,\,\,\,\,\int{\sqrt{-g}\left(\frac{1}{\sqrt{-g}}\frac{\partial(\sqrt{-g}\mathcal{L}_{m,r})}{\partial g^{\mu\nu}}\right)\delta g^{\mu\nu}d^4x},
\end{eqnarray}
where $f = f(R,G,T)$, $f_R=\frac{\partial f(R,G,T)}{\partial R}$, $f_G=\frac{\partial f(R,G,T)}{\partial G}$, and $f_T=\frac{\partial f(R,G,T)}{\partial T}$. Using $\delta S_{[g^{\mu\nu}]}=0$, eqs.(\ref{Xi}), and (\ref{metric var} -- \ref{T var}) one can gives,
\begin{eqnarray}
0&=&\frac{1}{2k^2} \int{\sqrt{-g}\left(f_R R_{\mu\nu} +f_R g_{\mu\nu}\square -f_R \nabla_\mu\nabla_\nu -\frac{1}{2} g_{\mu\nu} f\right)\delta g^{\mu\nu}d^4x}\nonumber\\
&+&\frac{1}{2k^2} \int{\sqrt{-g}\, \left(2R\left(f_GR_{\mu\nu}+f_Gg_{\mu\nu}\square -f_G \nabla_\mu\nabla_\nu\right)
-4f_G\left(R_{\mu}^{\rho}R_{\rho\nu}+R_{\mu\nu}\square+g_{\mu\nu}R^{\rho\lambda}\nabla_{\rho}\nabla_{\lambda}\right)\right)\delta g^{\mu\nu} d^{4}x}\nonumber\\
&-&\frac{4}{2k^2}\int{\sqrt{-g}\left(f_G\left(R_{\mu\rho\nu\lambda}R^{\rho\lambda}-\frac{1}{2}R_{\mu}^{\rho\lambda\xi}R_{\nu\rho\lambda\xi}\right)-f_G\left(R_{\mu}^{\rho}\nabla_{\mu}\nabla_{\rho}+R_{\nu}^{\rho}\nabla_{\mu}\nabla_{\rho}+R_{\mu\rho\nu\lambda}\nabla^{\rho}\nabla^{\lambda}\right) 
\right)\delta g^{\mu\nu}d^{4}x}\nonumber\\
&+&\frac{1}{2k^2} \int{\sqrt{-g}\left(f_T(T_{\mu\nu}^{(m,r)}+\Theta_{\mu\nu})\right)\delta g^{\mu\nu}d^4x}\nonumber\\
&+&\frac{1}{2\kappa^2_s}\int{\sqrt{-g}\left(\partial_{\mu}\phi\partial_{\nu}\phi-\partial_{\mu}\psi\partial_{\nu}\psi
-g_{\mu\nu}(\frac{1}{2}g^{\alpha\beta}\partial_{\alpha}\phi\partial_{\beta}\phi-\frac{1}{2}g^{\alpha\beta}\partial_{\alpha}\psi\partial_{\beta}\psi
-V(\phi,\psi))\right)\delta g^{\mu\nu}d^4x}\nonumber\\
&+&\frac{1}{2k^2}\int{\sqrt{-g}\left(\frac{2k^2}{\sqrt{-g}}\frac{\partial(\sqrt{-g}\mathcal{L}_{m,r})}{\partial g^{\mu\nu}}\right)\delta g^{\mu\nu}d^4x}.
\end{eqnarray}
Doing integration and neglecting the boundary contributions, we get:
\begin{eqnarray}
0&=& \int{\delta g^{\mu\nu}\left(f_R R_{\mu\nu}-\frac{1}{2} g_{\mu\nu} f + (g_{\mu\nu}\square - \nabla_\mu\nabla_\nu)f_R \right)d^4x}\nonumber\\
&+& \int{\delta g^{\mu\nu}\left(2R\left(f_GR_{\mu\nu}+(g_{\mu\nu}\square - \nabla_\mu\nabla_\nu)f_G\right)
-4\left(f_G R_{\mu}^{\rho}R_{\rho\nu}+(R_{\mu\nu}\square+g_{\mu\nu}R^{\rho\lambda}\nabla_{\rho}\nabla_{\lambda})f_G\right)\right) d^{4}x}\nonumber\\
&-&\int{\delta g^{\mu\nu}\left(f_G\left(4R_{\mu\rho\nu\lambda}R^{\rho\lambda}-2R_{\mu}^{\rho\lambda\xi}R_{\nu\rho\lambda\xi}\right)-4\left(R_{\mu}^{\rho}\nabla_{\mu}\nabla_{\rho}+R_{\nu}^{\rho}\nabla_{\mu}\nabla_{\rho}+R_{\mu\rho\nu\lambda}\nabla^{\rho}\nabla^{\lambda}\right)f_G 
\right)d^{4}x}\nonumber\\
&+& \int{\delta g^{\mu\nu}\left(-k^2 T_{\mu\nu}^{(T)}\right)d^4x}\nonumber\\
&+&\int{\delta g^{\mu\nu}\left(-\frac{k^2}{\kappa^2_s}T^{(\Xi)}_{\mu\nu}\right)d^4x}\nonumber\\
&+&\int{\delta g^{\mu\nu}\left(-k^2 T_{\mu\nu}^{(m,r)}\right)d^4x},
\end{eqnarray}
where
\begin{eqnarray}
T_{\mu\nu}^{(T)}&=&-f_T\left(\frac{T_{\mu\nu}^{(m,r)}+\Theta_{\mu\nu}}{k^2}\right),\\
T_{\mu\nu}^{(\Xi)}&=&-\partial_{\mu}\phi\partial_{\nu}\phi+\partial_{\mu}\psi\partial_{\nu}\psi
+\frac{1}{2}g_{\mu\nu}\left(g^{\alpha\beta}(\partial_{\alpha}\phi\partial_{\beta}\phi-\partial_{\alpha}\psi\partial_{\beta}\psi)
-2V(\phi,\psi)\right),\\
T_{\mu\nu}^{(m,r)}&=&-\frac{2}{\sqrt{-g}}\frac{\partial(\sqrt{-g}\mathcal{L}_{m,r})}{\partial g^{\mu\nu}}\cdot
\end{eqnarray}

\subsection{ Calculation of $\Theta_{\mu\nu}$}\label{subsec:Theta}
To calculate $\Theta_{\mu\nu}$, one can use equation (\ref{T(m)}) as,
\begin{eqnarray}
\Theta_{\mu\nu}&=&g^{\alpha\beta}\frac{\partial T_{\alpha\beta}^{(m,r)}}{\partial g^{\mu\nu}}=g^{\alpha\beta}\frac{\partial \left(g_{\alpha\beta}\mathcal{L}_{m,r}-2\frac{\partial\mathcal{L}_{m,r}}{\partial g^{\alpha\beta}}\right)}{\partial g^{\mu\nu}}=g^{\alpha\beta}\left(\frac{\partial g_{\alpha\beta}}{\partial g^{\mu\nu}}\mathcal{L}_{m,r}+g_{\alpha\beta}\frac{\partial \mathcal{L}_{m,r}}{\partial g^{\mu\nu}}-2\frac{\partial^2\mathcal{L}_{m,r}}{\partial g^{\alpha\beta}\partial g^{\mu\nu}}\right)\nonumber\\
&=&g^{\alpha\beta}\left(-g_{\alpha\sigma}g_{\beta\gamma}\delta^{\sigma\gamma}_{\mu\nu}\mathcal{L}_{m,r}+\frac{1}{2}g_{\alpha\beta}g_{\mu\nu}\mathcal{L}_{m,r}-\frac{1}{2}g_{\alpha\beta}T_{\mu\nu}^{(m,r)}-2\frac{\partial^2\mathcal{L}_{m,r}}{\partial g^{\alpha\beta}\partial g^{\mu\nu}}\right)\nonumber\\
&=&-g_{\mu\nu}\mathcal{L}_{m,r}+2g_{\mu\nu}\mathcal{L}_{m,r}-2T_{\mu\nu}^{(m,r)}-2g^{\alpha\beta}\frac{\partial^2\mathcal{L}_{m,r}}{\partial g^{\alpha\beta}\partial g^{\mu\nu}}\nonumber\\
&=&g_{\mu\nu}\mathcal{L}_{m,r}-2T_{\mu\nu}^{(m,r)}-2g^{\alpha\beta}\frac{\partial^2\mathcal{L}_{m,r}}{\partial g^{\alpha\beta}\partial g^{\mu\nu}}\cdot
\end{eqnarray}

\subsection{Covariant Divergence of Stress-Energy Tensor}\label{subsec:Covariant}
By using Eq.(\ref{T(R)}), we have
\begin{eqnarray}
\nabla^{\mu}T_{\mu\nu}^{(R)}&=&\nabla^{\mu}\left(f_R R_{\mu\nu}-\frac{1}{2} g_{\mu\nu} f + (g_{\mu\nu}\square - \nabla_\mu\nabla_\nu)f_R\right)\nonumber\\
&=& (\nabla^{\mu}f_R)R_{\mu\nu}+f_R\nabla^{\mu}R_{\mu\nu}-\frac{1}{2}f_R\nabla^{\mu}(g_{\mu\nu}R)-\frac{1}{2}f_G\nabla^{\mu}(g_{\mu\nu}G)-\frac{1}{2}f_T\nabla^{\mu}(g_{\mu\nu}T)\nonumber\\
&+&(\nabla_\nu\square-\square\nabla_\nu)f_R\nonumber\\
&=& R_{\mu\nu}\nabla^{\mu}f_R+f_R\nabla^{\mu}\left(R_{\mu\nu}-\frac{1}{2}g_{\mu\nu}R\right)-R_{\mu\nu}\nabla^{\mu}f_R-\frac{1}{2}f_G\nabla^{\mu}(g_{\mu\nu}G)-\frac{1}{2}f_T\nabla^{\mu}(g_{\mu\nu}T)\nonumber\\
&=&f_R\nabla^\mu G_{\mu\nu}-\frac{1}{2}f_G\nabla^{\mu}(g_{\mu\nu}G)-\frac{1}{2}f_T\nabla^{\mu}(g_{\mu\nu}T)\nonumber\\
&=&-\frac{1}{2}g_{\mu\nu}f_G\nabla^{\mu}G-\frac{1}{2}g_{\mu\nu}f_T\nabla^{\mu}T,\label{Co_T(R)1}
\end{eqnarray}
where $G_{\mu\nu}\doteq R_{\mu\nu}-\frac{1}{2}g_{\mu\nu}R$ is the Einstein tensor, and $\nabla^\mu G_{\mu\nu}=0$. 

By using Eq.(\ref{T(Xi)}), one can yield
\begin{eqnarray}
\nabla^{\mu}T_{\mu\nu}^{(\Xi)}=\dot{\phi}\left(\ddot{\phi}+3H\dot{\phi}-V_{,\phi}\right)-\dot{\psi}\left(\ddot{\psi}+3H\dot{\psi}+V_{,\psi}\right)\cdot
\end{eqnarray}
So, by comparing equations (\ref{EOM-phi}) and (\ref{EOM-psi}), we have
\begin{eqnarray}
\nabla^{\mu}T_{\mu\nu}^{(\Xi)}=0\cdot
\end{eqnarray}


\begin{thebibliography}{11}
%% 1 %%
\bibitem{khoury2002big}
J. Khoury et al. \textit{Phys. Rev. D}, 65(8):086007, (2002).
%% 2 %%
\bibitem{riess1998observational}
A. G. Riess et al. (Supernova Search Team Collabor ation), \textit{Astron. J.} 116: 1009, (1998);
%% 3 %%
\bibitem{perlmutter1999supernova}
S. Perlmutter et al. \textit{Astrophys. J.}  517: 565, (1999).
%% 4 %%
\bibitem{spergel2003first}
D. N. Spergel et al. \textit{Astrophys. J. Suppl. Ser.} 148: 175, (2003).
%% 5 %%
\bibitem{spergel2007three}
D. N. Spergel et al. \textit{Astrophys. J. Suppl. Ser.}, 170(2): 377, (2007).
%% 6 %%
\bibitem{Planck2018}
N. Aghanim et al. \textit{Astronomy \& Astrophys.}, 641: A6, (2023).
%% 7 %%
\bibitem{essen1990general}
H. Ess{\'e}n. \textit{Int. J.  The. Phys.}, 29(2):183--187, (1990).
%% 8 %%
\bibitem{Feynman1963Quantum}
R. Feynman. \textit{Acta Phys. Polon.} 24: 697, (1963).
%% 9 %%
\bibitem{fronsdal1984ghost}
C. Fronsdal. \textit{Phys. Rev. D}, 30(10): 2081, (1984).
%% 10 %%
\bibitem{faraoni1998conformal}
V. Faraoni, E. Gunzing and P. Nardone. \textit{Fund. Cosmic Phys.} 20: 121, (1999).
%% 11 %%
\bibitem{mannheim1997galactic}
P. D. Mannheim. \textit{Astrophys. J.}, 479(2): 659, (1997).
%% 12 %%
\bibitem{takook2010linear}
M. V. Takook and M. R. Tanhayi. \textit{JHEP}, 2010(12):1--15, (2010).
%% 13 %%
\bibitem{Feynman1995Feynman}
R. P. Feynman, N. L. F. B Morinigo, and W. G. Wagner. \textit{Addison-Wesley}, (1995).
%% 14 %%
\bibitem{farajollahi2012cosmological}
H. Farajollahi, F. Milani, et al. \textit{Astrophys.} $\&$ \textit{Spa. Sci.}, 337(2): 773--778, (2012).
%% 15 %%
\bibitem{farajollahi2011cosmic}
H. Farajollahi, F. Milani, et al. \textit{Gen. Rel.} $\&$ \textit{Grav.}, 43(6): 1657--1669, (2011).
%% 16 %%
\bibitem{Bamba2008Future}
K. Bamba, S. Nojiri and S. D. Odintsov. \textit{JCAP}, 0810: 045,F (2008).
%% 17 %%
\bibitem{capozziello2008extended}
S. Capozziello and M. Francaviglia. \textit{Gen. Rel. Grav.}, 40: 357--420, (2008).
%% 18 %%
\bibitem{sami2005fat}
M. Sami, A. Toporensky, P. V. Tretjakov and S. Tsujikawa. \textit{Phys. Lett. B}, 619: 193, (2005).
%% 19 %%
\bibitem{nojiri2005modified}
S. Nojiri and S. D. Odintsov. \textit{Phys. Lett. B}, 631: 1, (2005).
%% 20 %%
\bibitem{sadeghi2009crossing}
J. Sadeghi, M. R. Setare, A. Banijamali, and F. Milani. \textit{Phys. Rev. D}, 79(12): 123003, (2009).
%% 21 %%
\bibitem{atazadeh2014energy}
K. Atazadeh, F. Darabi. \textit{Gen. Rel. Grav.}, 46: 1--14, (2014).
%% 22 %%
\bibitem{bamba2010finite}
K. Bamba, S. D. Odintsov, L. Sebastiani and S. Zerbini. \textit{Eur. Phys. J. C} 67: 295 (2010).
%% 23 %%
\bibitem{Felice2010cosmological}
A. D. Felice, J. M. Gerard, T. Suyama. \textit{Phys. Rev. D} 82: 063526 (2010).
%% 24 %%
\bibitem{Felice2010inevitable}
A. De Felice, T. Tanaka. \textit{Prog. Theor. Phys.} 124: 503 (2010).
%% 25 %%
\bibitem{Shamir2017energy}
M. F. Shamir, and A. Komal. \textit{Int. J. Geo. Meth.  Mod. Phys.}, 14(12): 1750169 (2017).
%% 26 %%
\bibitem{Mustafa2020physically}
G. Mustafa,  M. F. Shamir, Xia Tie-Cheng. \textit{Phys. rev. D}, 101(10): 104013, (2020).
%% 27 %%
\bibitem{Navo2020stability}
G. Navo, and E. Elizalde. \textit{Int. J. Geo Meth. Mod. Phys.}, 17(11): 2050162m, (2020).
%% 28 %%
\bibitem{bhatti2022analysis}
M. Z. Bhatti, Z. Yousaf, A. Rehman. \textit{Int. J. Mod. Phys. D}, 31(01): 2150124, (2022).
%% 29 %%
\bibitem{Harko2011f(RT)gravity}
T. Harko, et al.  \textit{Phys. Rev. D}, 84(2): 024020,(2011).
%% 30 %%
\bibitem{baffou2021inflationary}
E. H. Baffou, et al. \textit{Annal. Phys.}, 434: 168620, (2021).
%% 31 %%
\bibitem{nashed2023theeffect}
G. G. L. Nashed. \textit{Astrophys. J.}, 950(2): 129, (2023).
%% 32 %%
\bibitem{singh2016cosmological}
G. P. Singh, B. K. Bishi, and P. K. Sahoo. \textit{Int. J. Geom. Meth. Mod. Phys.}, 13(05)
%% 33 %%
\bibitem{rajabi2017unimodular}
F. Rajabi, and K. Nozari. \textit{Phys. Rev. D}, 96(8): 084061, (2017).
%% 34 %%
\bibitem{singh2014reconstruction}
C. P. Singh, and V. Singh. \textit{Gen. Rel. Grav.}, 46: 1--14, (2014).
%% 35 %%
\bibitem{sharif2016energy}
M. Sharif, and A. Ikram. \textit{Eur. Phys.l J. C},  76: 1-13, (2016).
%% 36 %%
\bibitem{shamir2018gravastars}
M. F. Shamir, and M. Ahmad. \textit{Phys. Rev. D}, 97(10): 104031, (2018). 
%% 37 %%
\bibitem{shamir2021bouncing}
M. F. Shamir. \textit{Phys. D. Uni.}  32: 100794, (2021).
%% 38 %%
\bibitem{debnath2020constructions}
U. Debnath.  \textit{Int.l J. Mod. Phys. A}, 35(31): 2050203, (2020).
%% 39 %%
\bibitem{Chaudhary2023cosmological}
H. Chaudhary,  et al. \textit{Eur. Phys. J. C}, 83(10): 918, (2023).
%% 40 %%
\bibitem{ilyas2022energy}
M. Ilyas, et al. \textit{Phys. Scr.}, 98(1): 015016, (2022).
%% 41 %%
\bibitem{ilyas2024gravastars}
M. Ilyas. \textit{Canad. J. Phys.}, (2024).
%% 42 %%
\bibitem{ilyas2021compact}
M. Ilyas. \textit{Int. J. Mod. Phys. A} 36(24): 2150165, (2021).
%% 43 %%
\bibitem{elitzur2002big}
S. Elitzur et al. \textit{JHEP}, 2002(06): 017, (2002).
%% 44 %%
\bibitem{sadeghi2009bouncing}
J. Sadeghi, F. Milani, and A. R. Amani. \textit{Mod. Phys. Lett. A}, 24(29): 2363--2376, (2009).
%% 45 %%
\bibitem{bamba2014bounce}
K. Bamba et al. \textit{JCAP}, 2014(01): 008, (2014).
%% 46 %%
\bibitem{farajollahi2010bouncing}
H. Farajollahi and F. Milani. \textit{Mod. Phys. Lett. A}, 25(27): 2349--2362, (2010).
%% 47 %%
\bibitem{riess2004type}
A. G. Riess et al.  \textit{ApJ}, 607(2): 665, (2004).
%% 48 %%
\bibitem{knop2003new}
R. A. Knop et al. \textit{ApJ}, 598(1): 102, (2003).
%% 49 %%
\bibitem{riess1999bvri}
A. G. Riess et al. \textit{AsJ}, 117(2): 707, (1999).
%% 50 %%
\bibitem{perlmutter1999measurements}
S. Perlmutter et al. \textit{ApJ}, 517(2): 565, (1999).
%% 51 %%
\bibitem{ade2015planck}
P. A. Ade, et al.  \textit{Astron. \& Astrophys.}, 59: A144, (2016).
%% 52 %%
\bibitem{smecker1991type}
T. A. Smecker, and R. FG Wyse. \textit{AIP Con. Proc.}, 222(1): 467--470, (1991).
%% 53 %%
\bibitem{ruiz1995type}
P. Ruiz-Lapuente, A. Burkert, and R. Canal. \textit{Astrophys. J.}, 447(2): L69, (1995).
%% 54 %%
\bibitem{nomoto1997type}
K. Nomoto, K. Iwamoto, and N. Kishimoto. \textit{Science}, 276(5317): 1378-1382, (1997).
%% 55 %%
\bibitem{eisenstein2005detection}
D. J. Eisenstein et al. \textit{ApJ}, 633(2): 560, (2005).
%% 56 %%
\bibitem{astier2006supernova}
P. Astier et al. \textit{Astron.} $\&$ \textit{Astrophys.}, 447(1): 31--48, (2006).
%% 57 %%
\bibitem{riess2007new}
A. G. Riess et al. \textit{ApJ}, 659(1): 98, (2007).
%% 58 %%
\bibitem{wood2007observational}
W. M. Wood-Vasey et al. \textit{ApJ}, 666(2): 694, (2007).
%% 59 %%
\bibitem{bamba2012dark}
K. Bamba et al.  \textit{Astrophys.} $\&$ \textit{Spa. Sci.}, 342(1): 155--228, (2012).
%% 60 %%
\bibitem{padmanabhan2003cosmological}
T. Padmanabhan. \textit{Phys. Rep.}, 380(5-6): 235--320, (2003).
%% 61 %%
\bibitem{sahni2000case}
V. Sahni, and A. Starobinsky. \textit{Int. J. Mod. Phys. D}, 9(04): 373--443, (2000).
%% 62 %%
\bibitem{ratra1988cosmological}
B. Ratra, and P. JE Peebles. \textit{Phys. Rev. D}, 37(12): 3406, (1988).
%% 63 %%
\bibitem{caldwell2002phantom}
R. R Caldwell. \textit{Phys. Lett. B}, 545(1-2): 23--29, (2002).
%% 64 %%
\bibitem{feng2005dark}
B. Feng, X. Wang and X. Zhang. \textit{Phys. Lett. B}, 607(1):35--41, (2005).
%% 65 %%
\bibitem{guo2005cosmological}
Z. K. Guo, et al. \textit{Phys. Lett. B},  608(3-4): 177--182, (2005).
%% 66 %%
\bibitem{sen2002tachyon}
A. Sen. \textit{JHEP}, 2002(07): 065, (2002).
%% 67 %%
\bibitem{armendariz2000dynamical}
C. Armendariz-Picon, et al. \textit{Phys. Rev. Lett.}, 85(21): 4438, (2000).
%% 68 %%
\bibitem{gasperini2001quintessence}
M. Gasperini, F. Piazza, and G. Veneziano. \textit{Phys. Rev. D}, 65(2): 023508,(2001).
%% 69 %%
\bibitem{wei2005hessence}
H. Wei, R-G. Cai, and D-F. Zeng. \textit{Class. Quant. Grav.}, 22(16): 3189, (2005).
%% 70 %%
\bibitem{gumjudpai2009generalized}
B. Gumjudpai, and J. Ward. \textit{Phys. Rev. D}, 80(2): 023528, (2009).
%% 71 %%
\bibitem{martin2008dbi}
J. Martin, and M. Yamaguchi. \textit{Phys. Rev. D}, 77(12): 1235080, (2008).
%% 72 %%
\bibitem{bennett2003first}
C. L. Bennett et al. \textit{ApJS}, 148(1): 97, (2003).
%% 73 %%
\bibitem{bennett2011seven}
C. L. Bennett et al. \textit{ApJS}, 192(2): 17, (2011).
%% 74 %%
\bibitem{bennett20141}
C. L. Bennett et al. \textit{ApJ}, 794(2): 135, (2014).
%% 75 %%
\bibitem{bennett2013nine}
C. L. Bennett et al. \textit{ApJS}, 208(2): 20, (2013).
%% 76 %%
\bibitem{weinberg1989cosmological}
S. Weinberg. \textit{Rev. Mod. Phys.}, 61(1): 1, (1989).
%% 77 %%
\bibitem{tonry2003cosmological}
J. L. Tonry et al. \textit{ApJ}, 594(1): 1, (2003).
%% 78 %%
\bibitem{tegmark2004cosmological}
M. Tegmark et al. \textit{Phys. Rev. D},
69(10): 103501, (2004).
%% 79 %%
\bibitem{copeland2006dynamics}
E. J. Copeland, M. Sami, and S. Tsujikawa. \textit{Int. J. Mod. Phys. D},
15(11): 1753--1935, (2006).
%% 80 %%
\bibitem{seljak2005cosmological}
U. Seljak et al. \textit{Phys. Rev. D}, 71(10): 103515, (2005).
%% 81 %%
\bibitem{tegmark2005does}
M. Tegmark. \textit{JCAP}, 2005(04): 001, (2005).
%% 82 %%
\bibitem{ghanaatian2014bouncing}
M. Ghanaatian and F. Milani. \textit{Gen. Rel.} $\&$ \textit{Grav.}, 46(9): 1--16, (2014).
%% 83 %%
\bibitem{ghanaatian2018bouncing}
M. Ghanaatian, Mohammad, A. R. Gharaati, and F. Milani.  \textit{Annal. Phys.}, 397: 458--473, (2018).
%% 84 %%
\bibitem{sadeghi2008non}
J. Sadeghi, M. R. Setare, A. Banijamali, and F. Milani. 
\textit{Phys. Lett. B}, 662(2): 92, (2008).
%% 85 %%
\bibitem{farajollahi2011stability}
H. Farajollahi and F. Milani. \textit{IJTP}, 50(6):1953--1961, (2011).
%% 86 %%
\bibitem{farajollahi2010cosmic}
H. Farajollahi and A. Salehi. \textit{Int. J. Mod. Phy. D}, 19(05): 621--633, (2010).
%% 87 %%
\bibitem{farajollahi2011stability1}
H. Farajollahi and A. Salehi. \textit{JCAP}, 2011(07):036, (2011).
%% 88 %%
\bibitem{farajollahi2011stability2}
H. Farajollahi, A. Salehi, F. Tayebi, and A. Ravanpak. \textit{JCAP}, 2011(05): 017, (2011).
%% 89 %%
\bibitem{ron2003cosmic}
R. Cowen. \textit{Sci. News}, 163(10): 148-148, (2003).
%% 90 %%
\bibitem{schewe2003thebigrip}
R. F. Schewe. \textit{Phys. Today}, 56(10):9, (2003).
%% 91 %%
\bibitem{chimento2004onbigrip}
L. P. Chimento, and R. Lazkoz. \textit{Mod. Phys. Lett. A}, 19(33): 2479-2484, (2004).
%% 92 %%
\bibitem{caldwell2003phantom}
R. R. Caldwell, et al. \textit{Phys. Rev. Lett.}, 91(7): 071301, (2003).
%% 93 %%
\bibitem{carroll2003can}
S. M. Carroll, M. Hoffman, and M. Trodden. \textit{Phys. Rev. D}, 68(2): 023509, (2003).
%% 94 %%
\bibitem{cline2004phantom}
J. M. Cline, S. Jeon, and G. D. Moore. \textit{Phys. Rev. D}, 70(4):043543, (2004).
%% 95 %%
\bibitem{mcinnes2002ds}
B. McInnes. \textit{JHEP}, 2002(08):029, (2002).
%% 96 %%
\bibitem{melchiorri2003state}
A. Melchiorri, L. Mersini, C. J. \"{O}dman, and M. Trodden.  \textit{Phys. Rev. D}, 68(4):043509, (2003).
%% 97 %%
\bibitem{nesseris2004fate}
S. Nesseris and L. Perivolaropoulos. \textit{Phys. Rev. D}, 70(12): 123529, (2004).
%% 98 %%
\bibitem{onemli2004quantum}
V. K. Onemli and R. P. Woodard. \textit{Phys. Rev. D}, 70(10): 107301, (2004).
%% 99 %%
\bibitem{alam2004case}
U. Alam, V. Sahni, and A. A. Starobinsky. \textit{JCAP}, 2004(06):008, (2004).
%% 100 %%
\bibitem{padmanabhan2002accelerated}
T. Padmanabhan. \textit{Phys. Rev. D}, 66(2): 021301 1--021301 4, (2002).
%% 101 %%
\bibitem{padmanabhan2002can}
T. Padmanabhan and T. Roy Choudhury. \textit{Phys. Rev. D}, 66: 081301, (2002).
%% 102 %%
\bibitem{hao2003attractor}
J.-G. Hao and X.-Z. Li. \textit{Phys. Rev. D}, 67(10):107303,
(2003).
%% 103 %%
\bibitem{singh2003cosmological}
P. Singh, M. Sami, and N. Dadhich. \textit{Phys. Rev. D}, 68(2): 023522, (2003).
%% 104 %%
\bibitem{carroll2005can}
S. M. Carroll, A. De Felice, and M. Trodden. \textit{Phys. Rev. D}, 71(2): 023525, (2005).
%% 105 %%
\bibitem{aref2006exact}
I. Ya. Arefeva, S. Yu. Vernov, and A. S. Koshelev. \textit{Theo. Math. Phys.}, 148(1): 895, (2006).
%% 106 %%
\bibitem{guo2005interacting}
Z.-K. Guo and Y.-Z. Zhang. \textit{Phys. Rev. D}, 71(2):023501, (2005).
%% 107 %%
\bibitem{boisseau2000reconstruction}
B. Boisseau, et al. \textit{Phys. Rev. Lett.}, 85(11):2236, (2000).
%% 108 %%
\bibitem{sahni2003braneworld}
V. Sahni and Y. Shtanov. \textit{JCAP}, 2003(11): 014, (2003).
%% 109 %%
\bibitem{mcinnes2005phantom}
B. McInnes. \textit{Nucl. Phys. B}, 718(1): 55--82, (2005).
%% 110 %%
\bibitem{perivolaropoulos2005constraints}
L. Perivolaropoulos. \textit{Phys. Rev. D}, 71(6): 063503, (2005).
%% 111 %%
\bibitem{anisimov2005b}
A. Anisimov, E. Babichev, and A. Vikman. \textit{JCAP}, 2005(06): 006, (2005).
%% 112 %%
\bibitem{witten1986non}
E. Witten. \textit{Nucl. Phys. B}, 268(2): 253--294,
(1986).
%% 113 %%
\bibitem{witten1986interacting}
E. Witten. \textit{ Nucl. Phys. B}, 276(2): 291--324, (1986).
%% 114 %%
\bibitem{Guth1981}
A. H. Guth. \textit{Phys. Rev. D}, 23: 347-356, (1981).
%% 115 %%
\bibitem{Linde2008}
A. Linde. \textit{Contemporary Phys.}, 49: 552-567, (2008).
%% 116 %%
\bibitem{Peiris2003}
V. H. Peiris, et. al. \textit{Astrophys. J. Suppl. Ser.}, 148: 213-231, (2003).
%% 117 %%
\bibitem{Vilenkin1983}
A. Vilenkin, \textit{Phys. Reports}, 121(5-6): 263--270, (1985).
%% 118 %%
\bibitem{mukhanov2005physical}
V. F. Mukhanov. \textit{Camb. uni. pr.}, (2005).
\end{thebibliography}
\end{document}